\documentclass[12pt]{article}
\usepackage{amsmath,amssymb}
\usepackage{graphicx}
\newcommand{\eqdef}{\stackrel{\text{def}}{=}}
\newcommand{\n}{\nonumber\\}
\newcommand{\bm}{\boldsymbol}
\newcommand{\ignore}[1]{}
\numberwithin{equation}{section}
\newcommand{\Romannumeral}[1]{\uppercase\expandafter{\romannumeral#1}}
\newcommand{\I}{\text{\Romannumeral{1}}}
\newcommand{\II}{\text{\Romannumeral{2}}}

\newtheorem{prop}{\bf Proposition}%[section]

\allowdisplaybreaks[4]

\begin{document}

\baselineskip=20pt

%%%%%%%%%%%%%%%%%%%%%%%%%%%%%%%%%%%%%%%%%%%%%%%%%%%%%%%%%%%%
%                                                          %
%  Title page                                              %
%                                                          %
%%%%%%%%%%%%%%%%%%%%%%%%%%%%%%%%%%%%%%%%%%%%%%%%%%%%%%%%%%%%
\newcommand{\preprint}{
   \begin{flushright}\normalsize \sf
     DPSU-13-2\\
%     {\tt arXiv:1308.4240[math-ph]}\\
%     August 2013
   \end{flushright}}
\newcommand{\Title}[1]{{\baselineskip=26pt
   \begin{center} \Large \bf #1 \\ \ \\ \end{center}}}
\newcommand{\Author}{\begin{center}
   \large \bf Satoru Odake${}^a$ and Ryu Sasaki${}^{a,b}$ \end{center}}
\newcommand{\Address}{\begin{center}
     $^a$ Department of Physics, Shinshu University,\\
     Matsumoto 390-8621, Japan\\
     ${}^b$ Center for Theoretical Sciences,\\
    National Taiwan University, Taipei 10617, Taiwan
       \end{center}}
\newcommand{\Accepted}[1]{\begin{center}
   {\large \sf #1}\\ \vspace{1mm}{\small \sf Accepted for Publication}
   \end{center}}

\preprint
\thispagestyle{empty}

\Title{Casoratian Identities\\
for the Wilson and Askey-Wilson Polynomials
\footnote[5]{Dedicated to Richard Askey for his eightieth birthday.}
}
\Author

\Address
\vspace{1cm}

\begin{abstract}
Infinitely many Casoratian identities are derived for the Wilson and
Askey-Wilson polynomials in parallel to the Wronskian identities for the
Hermite, Laguerre and Jacobi polynomials, which were reported recently by
the present authors. These identities form the basis of the equivalence
between eigenstate adding and deleting Darboux transformations for solvable
(discrete) quantum mechanical systems. Similar identities hold for various
reduced form polynomials of the Wilson and Askey-Wilson polynomials,
{\em e.g.} the continuous $q$-Jacobi, continuous (dual) ($q$-)Hahn,
Meixner-Pollaczek, Al-Salam-Chihara, continuous (big) $q$-Hermite, etc.
\end{abstract}

%%%%%%%%%%%%%%%%%%%%%%%%%%%%%%%%%%%%%%%%%%%%%%%%%%%%%%%%%%%%%%%
%                                                             %
%  1. Introduction                                            %
%                                                             %
%%%%%%%%%%%%%%%%%%%%%%%%%%%%%%%%%%%%%%%%%%%%%%%%%%%%%%%%%%%%%%%
\section{Introduction}
\label{intro}

In a previous paper \cite{os29} we reported  infinitely many Wronskian
identities for the Hermite, Laguerre and Jacobi polynomials.
They relate the Wronskians of polynomials of {\em twisted\/} parameters to
the Wronskians of polynomials of {\em shifted\/} parameters.
Here we will present similar identities for the Wilson and Askey-Wilson
polynomials  and their reduced form polynomials \cite{askey,ismail,koeswart}.
The Wronskians are now replaced by their difference analogues, the Casoratians.

The basic logic of deriving these identities is the same for the Jacobi
polynomials etc and for the Askey-Wilson polynomials etc;
the equivalence between the multiple Darboux-Crum transformations
\cite{darb}--\cite{gos} in terms of {\em pseudo virtual state wave functions\/}
and those in terms of {\em eigenfunctions with shifted parameters\/}.
In other words, the duality between eigenstates adding and deleting
transformations. The virtual and pseudo virtual state wave functions have
been reported in detail for the differential and difference Schr\"odinger
equations \cite{os29,os25,os28,os27,os26}.
The virtual state wave functions are the essential ingredient for constructing 
multi-indexed orthogonal polynomials. The pseudo virtual state wave functions
play the main role in the above mentioned duality.
These Casoratian (Wronskian) identities could be understood as the
consequences of the {\em forward and backward shift relations\/} and the
{\em discrete symmetries\/} of the governing Schr\"odinger equations.
The forward and backward shift relations are the characteristic properties
of the {\em classical orthogonal polynomials\/}, satisfying second order
differential and difference equations.
These polynomials depend on a set of parameters, to be denoted symbolically
by $\bm{\lambda}$.
The forward shift operator $\mathcal{F}(\bm{\lambda})$ connects
$\check{P}_n(x;\bm{\lambda})$ to $\check{P}_{n-1}(x;\bm{\lambda}+\bm{\delta})$,
with $\bm{\delta}$ being the shift of the parameters.
For the definition of $\check{P}_n(x;\bm{\lambda})$, see \eqref{difSch} and
the paragraph below it.
The backward shift operator $\mathcal{B}(\bm{\lambda})$ connects them in
the opposite direction, see \eqref{FP=,BP=}.
In the context of quantum mechanical reformulation of the classical
orthogonal polynomials \cite{os24}, the principle underlying the forward
and backward shift relations is called {\em shape invariance\/} \cite{genden}.

These identities imply the equality of the deformed potential functions with the
twisted and shifted parameters in the difference Schr\"odinger equations.
This in turn guarantees the equivalence of all the other eigenstate
wave functions for proper parameter ranges if the self-adjointness of the
deformed Hamiltonian and  other requirements of quantum mechanical formulation
are satisfied.
In contrast, the Casoratian identities \eqref{poldual1}--\eqref{poldual2},
\eqref{detiden}--\eqref{genwronide} are purely algebraic relations and
they are valid at generic values of the parameters.

The present work is most closely related in its contents with \cite{os27},
which formulates deformations of the Wilson and Askey-Wilson polynomials
through Casoratians of virtual state wave functions.
The relationship of the present work with \cite{os27} is the same as that of
\cite{os29} with \cite{os25,os28}; derivation of Wronskian-Casoratian identities
which reflect the solvability of classical orthogonal polynomials revealed
through deformations.

This paper is organised as follows.
The formulation of the Wilson and Askey-Wilson polynomials through the
difference Schr\"odinger equations is recapitulated in section two.
The basic formulas of these polynomials necessary for the present purposes
are summarised in \S\,\ref{sec:basic}. The pseudo virtual states for the
Wilson and Askey-Wilson polynomials are introduced and discussed in
\S\,\ref{sec:pseudovs}.
Starting with the general properties the Casoratian determinants
in \S\,\ref{sec:casform}, the eigenstates adding Darboux transformations are
recapitulated in \S\,\ref{sec:adddarb}. The eigenstates deleting Darboux
transformations are summarised in \S\,\ref{sec:deldarb}. The Casoratian
identities for the Wilson and Askey-Wilson polynomials are presented in
\S\,\ref{sec:deriden}. This is the main part of the paper.
In section four the Casoratian identities are discussed for the other
classical orthogonal polynomials which are obtained by reductions from
the Wilson and Askey-Wilson polynomials. The basic formulas of the reduced
polynomials are summarised in sections \S\,\ref{sec:redW} and
\S\,\ref{sec:redAW}. The pseudo virtual state wave functions for the reduced
cases are introduced in \S\,\ref{sec:pseudoW}, \S\,\ref{sec:pseudoA} and
\S\,\ref{sec:pseudoB}. The Casoratian identities for the reduced polynomials
are discussed in \S\,\ref{sec:redCas}.
The final section is for a summary and comments.

%%%%%%%%%%%%%%%%%%%%%%%%%%%%%%%%%%%%%%%%%%%%%%%%%%%%%%%%%%%%%%%
%                                                             %
%  2. Pseudo Virtual States in Discrete Quantum Mechanics     %
%                                                             %
%%%%%%%%%%%%%%%%%%%%%%%%%%%%%%%%%%%%%%%%%%%%%%%%%%%%%%%%%%%%%%%
\section{Pseudo Virtual States in Discrete Quantum Mechanics }
\label{sec:pseudo}

Various properties of the classical orthogonal polynomials can be understood
in a unified fashion by considering them as the main part of the
eigenfunctions of a certain self-adjoint operator (called the Hamiltonian
or the Schr\"odinger operator) acting on a Hilbert space.
This scheme works for those classical orthogonal polynomials satisfying
second order difference equations (with real or pure imaginary shifts,
{\em e.g.} the Askey-Wilson \cite{os13} and $q$-Racah polynomials \cite{os12})
as well as for those obeying second order differential equations, {\em e.g.}
the Jacobi polynomials.
We refer to \cite{os24} for the general introduction of the quantum mechanical
reformulation of the classical orthogonal polynomials.

Here we first summarise the basic structure of discrete quantum mechanics
with pure imaginary shifts in one dimension.
Next in \S\,\ref{sec:pseudovs} we introduce the pseudo virtual state wave
functions, the key ingredient of the eigenstates adding transformations.
The general definitions and formulas are followed by explicit ones for
the Wilson and Askey-Wilson polynomials, which are two most generic members
of Askey scheme of hypergeometric orthogonal polynomials with pure imaginary
shifts.

%%%%%%%%%%%%%%%%%%%%%%%%%%%%%%%%%%%%%%%%%%%%%
%                                           %
% 2.1 Basic formulation                     %
%                                           %
%%%%%%%%%%%%%%%%%%%%%%%%%%%%%%%%%%%%%%%%%%%%%
\subsection{Basic formulation}
\label{sec:basic}

Here we summarise the basic definitions and formulas of discrete quantum
mechanics, with  the Wilson  and Askey-Wilson polynomials as explicit examples.
We start from the following factorised positive semi-definite Hamiltonian
$\bigl((e^{\pm\gamma p}f)(x)=f(x\mp i\gamma)\bigr)$:
\begin{align}
  &\mathcal{H}(\bm{\lambda})\eqdef
  \sqrt{V(x;\bm{\lambda})}\,e^{\gamma p}\sqrt{V^*(x;\bm{\lambda})}
  +\!\sqrt{V^*(x;\bm{\lambda})}\,e^{-\gamma p}\sqrt{V(x;\bm{\lambda})}
  -V(x;\bm{\lambda})-V^*(x;\bm{\lambda})
  \label{H}\\
  &\phantom{\mathcal{H}(\bm{\lambda})}
  =\mathcal{A(\bm{\lambda})}^{\dagger}\mathcal{A}(\bm{\lambda}),\n
  &\mathcal{A}(\bm{\lambda})\eqdef
  i\bigl(e^{\frac{\gamma}{2}p}\sqrt{V^*(x; \bm{\lambda})}
  -e^{-\frac{\gamma}{2}p}\sqrt{V(x;\bm{\lambda})}\,\bigr),\n
  &\mathcal{A}(\bm{\lambda})^{\dagger}\eqdef
  -i\bigl(\sqrt{V(x;\bm{\lambda})}\,e^{\frac{\gamma}{2}p}
  -\sqrt{V^*(x;\bm{\lambda})}\,e^{-\frac{\gamma}{2}p}\bigr),
\end{align}
which is an analytic difference operator acting on holomorphic functions
of $x$ on a strip, $x_1<\text{Re}\,x<x_2$, ($x_1,x_2\in\mathbb{R}$).
Here $p=-i\partial_x$ is the momentum operator and $\gamma$ is a real number.
The $*$-operation on an analytic function $f(x)=\sum_na_nx^n$
($a_n\in\mathbb{C}$) is defined by $f^*(x)=f(x^*)^*=\sum_na_n^*x^n$, in which
$a_n^*$ is the complex conjugation of $a_n$.
Obviously $f^{**}(x)=f(x)$.
If a function satisfies $f^*=f$, then it takes real values on the real line.
For the concrete forms of $V(x)$, see \eqref{Vform}.
The branch of $\sqrt{V(x)}$ is determined by the requirement of the
self-adjointness of the Hamiltonian \cite{os24,os13}.

The following type of factorisation of the eigenfunctions is characteristic
to all the systems related
with the classical orthogonal polynomials \cite{os24}, {\em e.g.\/} Jacobi
\cite{os25}, Askey-Wilson \cite{os27} and $q$-Racah \cite{os12}:
\begin{equation}
  \mathcal{H}(\bm{\lambda})\phi_n(x;\bm{\lambda})
  =\mathcal{E}_n(\bm{\lambda})\phi_n(x;\bm{\lambda}),\quad
  \phi_n(x;\bm{\lambda})
  =\phi_0(x;\bm{\lambda})\check{P}_n(x;\bm{\lambda})
  \ \ (n=0,1,2,\ldots),
  \label{difSch}
\end{equation}
in which $\phi_0(x;\bm{\lambda})$ is the ground state eigenfunction and
$\check{P}_n(x;\bm{\lambda})=P_n\bigl(\eta(x);\bm{\lambda}\bigr)$ is a
polynomial of degree $n$ in a certain function $\eta(x)$,
called the sinusoidal coordinate \eqref{etadef} \cite{os7}.
We adopt the convention of `real' eigenfunctions,
$\phi^*_0(x;\bm{\lambda})=\phi_0(x;\bm{\lambda})$ and
$\check{P}^*_n(x;\bm{\lambda})=\check{P}_n(x;\bm{\lambda})$.
The eigenfunctions form an orthogonal basis
\begin{align}
  (\phi_n,\phi_m)&\eqdef
  \int_{x_1}^{x_2}\!\!dx\,\phi_n^*(x;\bm{\lambda})\phi_m(x;\bm{\lambda})=
  \int_{x_1}^{x_2}\!\!dx\,\phi_0(x;\bm{\lambda})^2
  \check{P}_n(x;\bm{\lambda})\check{P}_m(x;\bm{\lambda})\n
  &=h_n(\bm{\lambda})\delta_{nm}
  \quad(n,m=0,1,2,\ldots),\quad 0<h_n(\bm{\lambda})<\infty.
  \label{ortrel}
\end{align}
The defining domain and the parameters for the Wilson (W) and Askey-Wilson
(AW) polynomials are:
\begin{alignat}{2}
  \text{W}:\ \ &x_1=0,\ x_2=\infty,\ \gamma=1,
  &\ \ \bm{\lambda}=(a_1,a_2,a_3,a_4),
  &\ \ \bm{\delta}=(\tfrac12,\tfrac12,\tfrac12,\tfrac12),
  \ \ \kappa=1,\n
  \text{AW}:\ \ &x_1=0,\ x_2=\pi,\ \gamma=\log q,
  &\ \ q^{\bm{\lambda}}=(a_1,a_2,a_3,a_4),
  &\ \ \bm{\delta}=(\tfrac12,\tfrac12,\tfrac12,\tfrac12),
  \ \ \kappa=q^{-1},
\end{alignat}
where $q^{\bm{\lambda}}$ stands for
$q^{(\lambda_1,\lambda_2,\ldots)}=(q^{\lambda_1},q^{\lambda_2},\ldots)$
and $0<q<1$. 
Here $\bm{\delta}$ is the shift of the parameters, which appears in various
relations, for example, \eqref{Fdef}--\eqref{FP=,BP=} and
\eqref{Vshape}--\eqref{phi0shape3} and $\kappa$ is a multiplicative constant 
of the potential function $V$ and others which appears in various formulas,
{\em e.g.\/} \eqref{Vshape}, \eqref{VXi},
\eqref{enformula1}-\eqref{enformula2}, \eqref{xiDn1}-\eqref{xiDn2}.
The parameters are restricted by
\begin{equation}
  \{a_1^*,a_2^*,a_3^*,a_4^*\}=\{a_1,a_2,a_3,a_4\}\ \ (\text{as a set});\quad
  \text{W}:\ \text{Re}\,a_i>0,\quad
  \text{AW}:\ |a_i|<1.
  \label{rangeorg}
\end{equation}
Here are the fundamental data:
\begin{align}
  &V(x;\bm{\lambda})=\left\{
  \begin{array}{ll}
  \bigl(2ix(2ix+1)\bigr)^{-1}\prod_{j=1}^4(a_j+ix)
  &:\text{W}\\[4pt]
  \bigl((1-e^{2ix})(1-qe^{2ix})\bigr)^{-1}\prod_{j=1}^4(1-a_je^{ix})
  &:\text{AW}
  \end{array}\right.,
  \label{Vform}\\[2pt]
  &\eta(x)=\left\{
  \begin{array}{ll}
  x^2&:\text{W}\\
  \cos x&:\text{AW}
  \end{array}\right.,\quad
  \varphi(x)=\left\{
  \begin{array}{ll}
  2x&:\text{W}\\
  2\sin x&:\text{AW}
  \end{array}\right.,
  \label{etadef}\\
  &\mathcal{E}_n(\bm{\lambda})=\left\{
  \begin{array}{lll}
  n(n+b_1-1),&b_1\eqdef a_1+a_2+a_3+a_4&:\text{W}\\[2pt]
  (q^{-n}-1)(1-b_4q^{n-1}),&b_4\eqdef a_1a_2a_3a_4&:\text{AW}
  \end{array}\right.,
  \label{b1b4}\\
  &\phi_n(x;\bm{\lambda})
  =\phi_0(x;\bm{\lambda})\check{P}_n(x;\bm{\lambda}),
  \label{factphin}\\
  &\check{P}_n(x;\bm{\lambda})=P_n\bigl(\eta(x);\bm{\lambda}\bigr)
  =\left\{\begin{array}{ll}
  W_n\bigl(\eta(x);a_1,a_2,a_3,a_4\bigr)&:\text{W}\\[2pt]
  p_n\bigl(\eta(x);a_1,a_2,a_3,a_4|q\bigr)&:\text{AW}
  \end{array}\right.\n
  &\phantom{\check{P}_n(x;\bm{\lambda})}=\left\{\begin{array}{ll}
  {\displaystyle
  (a_1+a_2,a_1+a_3,a_1+a_4)_n}\\[2pt]
  {\displaystyle
  \quad\times
  {}_4F_3\Bigl(
  \genfrac{}{}{0pt}{}{-n,\,n+b_1-1,\,a_1+ix,\,a_1-ix}
  {a_1+a_2,\,a_1+a_3,\,a_1+a_4}\Bigm|1\Bigr)
  }&:\text{W}\\[8pt]
  {\displaystyle
  a_1^{-n}(a_1a_2,a_1a_3,a_1a_4\,;q)_n}\\[2pt]
  {\displaystyle
  \quad\times
  {}_4\phi_3\Bigl(\genfrac{}{}{0pt}{}{q^{-n},\,b_4q^{n-1},\,
  a_1e^{ix},\,a_1e^{-ix}}{a_1a_2,\,a_1a_3,\,a_1a_4}\!\!\Bigm|\!q\,;q\Bigr)
  }&:\text{AW}
  \end{array}\right.,
  \label{Pn=W,AW}\\[2pt]
  &\phi_0(x;\bm{\lambda})=\left\{
  \begin{array}{ll}
  \sqrt{\bigl(\Gamma(2ix)\Gamma(-2ix)\bigr)^{-1}\prod_{j=1}^4
  \Gamma(a_j+ix)\Gamma(a_j-ix)}&:\text{W}\\[5pt]
  \sqrt{(e^{2ix},e^{-2ix}\,;q)_{\infty}
  \prod_{j=1}^4(a_je^{ix},a_je^{-ix}\,;q)_{\infty}^{-1}}
  &:\text{AW}
  \end{array}\right..
  \label{phi0=W,AW}
\end{align}
Here $W_n$ and $p_n$ in \eqref{Pn=W,AW} are the Wilson and the Askey-Wilson
polynomials defined in \cite{koeswart} and the symbols $(a)_n$ and $(a;q)_n$
are ($q$-)shifted factorials.
The auxiliary function $\varphi(x)$ \eqref{etadef} connects
$V(x;\bm{\lambda}+\bm{\delta})$ with $V(x-i\tfrac{\gamma}{2};\bm{\lambda})$
\eqref{Vshape} and $\phi_0(x;\bm{\lambda}+\bm{\delta})$ with
$\phi_0(x+i\tfrac{\gamma}{2};\bm{\lambda})$ \eqref{phi0shape} and others.

The most basic ingredient of this formulation is the {\em ground state
eigenfunction\/} $\phi_0(x;\bm{\lambda})$, which is the {\em zero mode}
of the operator $\mathcal{A}(\bm{\lambda})$:
\begin{equation}
  \mathcal{A}(\bm{\lambda})\phi_0(x;\bm{\lambda})=0
  \ \,\Rightarrow
  \sqrt{V^*(x-i\tfrac{\gamma}{2};\bm{\lambda})}
  \,\phi_0(x-i\tfrac{\gamma}{2};\bm{\lambda})
  =\sqrt{V(x+i\tfrac{\gamma}{2};\bm{\lambda})}
  \,\phi_0(x+i\tfrac{\gamma}{2};\bm{\lambda}).
  \label{zero}
\end{equation}
The essential property of the ground state wave function
$\phi_0(x;\bm{\lambda})$ \eqref{zero} is that it has no zeros in the domain
$x_1<x<x_2$, and its square gives the weight function of the classical
orthogonal polynomials \eqref{ortrel}. In other words, the quantum mechanical
reformulation provides the weight functions of the classical orthogonal
polynomials based only on the data ($V(x)$) of the difference equation of
the polynomials \eqref{Htil}, \eqref{HtP=EP}.
The situation is the same for the ($q$-)Racah polynomials, etc \cite{os12}.
This reformulation, in turn, opens various possibilities for deformations.
By similarity transforming the difference Schr\"odinger equation
\eqref{difSch} in terms of the ground state eigenfunction, we obtain the
second order difference operator $\widetilde{\mathcal{H}}(\bm{\lambda})$
acting on the polynomial eigenfunctions
\begin{align}
  &\widetilde{\mathcal{H}}(\bm{\lambda})\eqdef
  \phi_0(x;\bm{\lambda})^{-1}\circ\mathcal{H}(\bm{\lambda})
  \circ\phi_0(x;\bm{\lambda})\n
  &\phantom{\widetilde{\mathcal{H}}_{\ell}(\bm{\lambda})}
  =V(x;\bm{\lambda})(e^{\gamma p}-1)
  +V^*(x;\bm{\lambda})(e^{-\gamma p}-1),
  \label{Htil}\\
  &\widetilde{\mathcal{H}}(\bm{\lambda})\check{P}_n(x;\bm{\lambda})
  =\mathcal{E}_n(\bm{\lambda})\check{P}_n(x;\bm{\lambda}),
  \label{HtP=EP}
\end{align}
and $\widetilde{\mathcal{H}}(\bm{\lambda})=\mathcal{B}(\bm{\lambda})
\mathcal{F}(\bm{\lambda})$ is {\em square root free}.
This is the conventional difference equation for the Wilson and Askey-Wilson
polynomials and their reduced form polynomials.
The forward and backward shift operators $\mathcal{F}(\bm{\lambda})$ and
$\mathcal{B}(\bm{\lambda})$, which express the shape invariance relations,
are defined by
\begin{align}
  &\mathcal{F}(\bm{\lambda})\eqdef
  \phi_0(x;\bm{\lambda}+\bm{\delta})^{-1}\circ
  \mathcal{A}(\bm{\lambda})\circ\phi_0(x;\bm{\lambda})
  =i\varphi(x)^{-1}(e^{\frac{\gamma}{2}p}-e^{-\frac{\gamma}{2}p}),
  \label{Fdef}\\
  &\mathcal{B}(\bm{\lambda})\eqdef
  \phi_0(x;\bm{\lambda})^{-1}\circ
  \mathcal{A}(\bm{\lambda})^{\dagger}
  \circ\phi_0(x;\bm{\lambda}+\bm{\delta})
  =-i\bigl(V(x;\bm{\lambda})e^{\frac{\gamma}{2}p}
  -V^*(x;\bm{\lambda})e^{-\frac{\gamma}{2}p}\bigr)\varphi(x),
  \label{Bdef}
\end{align}
and their action on the polynomials is
\begin{equation}
  \mathcal{F}(\bm{\lambda})\check{P}_n(x;\bm{\lambda})
  =f_n(\bm{\lambda})\check{P}_{n-1}(x;\bm{\lambda}+\bm{\delta}),\quad
  \mathcal{B}(\bm{\lambda})\check{P}_{n-1}(x;\bm{\lambda}+\bm{\delta})
  =b_{n-1}(\bm{\lambda})\check{P}_n(x;\bm{\lambda}).
  \label{FP=,BP=}
\end{equation}
These are universal relations valid for all the polynomials in the Askey
scheme. In the above equations, the factors of the energy eigenvalue,
$f_n(\bm{\lambda})$ and $b_{n-1}(\bm{\lambda})$,
$\mathcal{E}_n(\bm{\lambda})=f_n(\bm{\lambda})b_{n-1}(\bm{\lambda})$,
for the Wilson and Askey-Wilson polynomials are given by
\begin{equation}
  f_n(\bm{\lambda})=\left\{
  \begin{array}{ll}
  -n(n+b_1-1)&:\text{W}\\
  q^{\frac{n}{2}}(q^{-n}-1)(1-b_4q^{n-1})&:\text{AW}
  \end{array}\right.,
  \quad
  b_{n-1}(\bm{\lambda})=\left\{
  \begin{array}{ll}
  -1&:\text{W}\\
  q^{-\frac{n}{2}}&:\text{AW}
  \end{array}\right.,
\end{equation}
and the auxiliary function $\varphi(x)$ is defined in \eqref{etadef}.
Here $b_{n-1}({\bm{\lambda}})$ given above should not be confused with
$b_1$ and $b_4$ as given in \eqref{b1b4}.

At the basis of these relations are the shape covariant properties of the
potential and the ground state eigenfunctions \cite{os27}:
\begin{align}
  &V(x;\bm{\lambda}+\bm{\delta})
  =\kappa^{-1}\frac{\varphi(x-i\gamma)}{\varphi(x)}
  V(x-i\tfrac{\gamma}{2};\bm{\lambda}),
  \label{Vshape}\\
  &\phi_0(x;\bm{\lambda}+\bm{\delta})
  =\varphi(x)\sqrt{V(x+i\tfrac{\gamma}{2};\bm{\lambda})}\,
  \phi_0(x+i\tfrac{\gamma}{2};\bm{\lambda}),
  \label{phi0shape}\\
  &\Bigl(\Rightarrow\ \ \phi_0(x;\bm{\lambda})
  =\varphi(x)\sqrt{V(x+i\tfrac{\gamma}{2};\bm{\lambda}-\bm{\delta})}\,
  \phi_0(x+i\tfrac{\gamma}{2};\bm{\lambda}-\bm{\delta})
   \label{phi0shape2}\\
  &\phantom{\Bigl(\Rightarrow\ \ \phi_0(x;\bm{\lambda})}
  =\varphi(x)\sqrt{V^*(x-i\tfrac{\gamma}{2};\bm{\lambda}-\bm{\delta})}\,
  \phi_0(x-i\tfrac{\gamma}{2};\bm{\lambda}-\bm{\delta})\ \Bigr).
  \label{phi0shape3}
\end{align}

For the purpose of rational extensions of these classical orthogonal
polynomials, deformations of difference Schr\"odinger equations
\eqref{H}--\eqref{difSch} have proved fruitful, rather than those of the
above difference equations \eqref{Htil}--\eqref{HtP=EP}.
The analogue of multiple Darboux transformations for the difference
Schr\"odinger equations \eqref{H}--\eqref{difSch} had been formulated by
the present authors some years ago \cite{os15,gos}.
By choosing special types of non-eigen {\em seed solutions\/}, called the
{\em virtual state wave functions\/} \cite{os24}, the {\em multi-indexed\/}
Wilson and Askey-Wilson polynomials had been constructed \cite{os27}.
In those cases, the deformed systems are exactly iso-spectral to the
original system.

In the present paper, we consider non-isospectral deformations by using
the {\em pseudo virtual state wave functions\/} \cite{os29,os28} as in the
parallel situations for the Jacobi polynomials etc. \cite{os29}.

%%%%%%%%%%%%%%%%%%%%%%%%%%%%%%%%%%%%%%%%%%%%%
%                                           %
% 2.2 Pseudo virtual state wave functions   %
%                                           %
%%%%%%%%%%%%%%%%%%%%%%%%%%%%%%%%%%%%%%%%%%%%%
\subsection{Pseudo virtual state wave functions}
\label{sec:pseudovs}

The {\em pseudo virtual state wave functions} are defined from the
eigenfunctions by {\em twisting\/} the parameters,
$\bm{\lambda}\to\mathfrak{t}(\bm{\lambda})$, $ \mathfrak{t}^2=\text{Id}$,
based on the discrete symmetry of the original Hamiltonian system \eqref{H}.

For a certain choice of the twist operator $\mathfrak{t}$, the twisted
potential function $V'(x;\bm{\lambda})$
\begin{equation}
  V'(x;\bm{\lambda})\eqdef V\bigl(x;\mathfrak{t}(\bm{\lambda})\bigr),
  \label{V'def1}
\end{equation}
satisfies the relations
\begin{align}
  V(x;\bm{\lambda})V^*(x-i\gamma;\bm{\lambda})
  &=\alpha(\bm{\lambda})^2V'(x;\bm{\lambda})V^{\prime*}(x-i\gamma;\bm{\lambda}),
  \label{VV'rel}\\
  V(x;\bm{\lambda})+V^*(x;\bm{\lambda})
  &=\alpha(\bm{\lambda})\bigl(V'(x;\bm{\lambda})
  +V^{\prime*}(x;\bm{\lambda})\bigr)-\alpha'(\bm{\lambda}),
  \label{propV'}
\end{align}
with real constants  $\alpha(\bm{\lambda})$ and $\alpha'(\bm{\lambda})$.
The second condition \eqref{propV'} determines the sign of
$\alpha(\bm{\lambda})$.
These mean a linear relation between the two Hamiltonians:
\begin{align}
  \mathcal{H}(\bm{\lambda})
  &=\alpha(\bm{\lambda})\mathcal{H}'(\bm{\lambda})+\alpha'(\bm{\lambda}),
  \label{H=aH'+a'}\\
  \mathcal{H}'(\bm{\lambda})
  &\eqdef\sqrt{V'(x;\bm{\lambda})}\,e^{\gamma p}
  \sqrt{V^{\prime*}(x;\bm{\lambda})}
  +\!\sqrt{V^{\prime*}(x;\bm{\lambda})}\,e^{-\gamma p}
  \sqrt{V'(x;\bm{\lambda})}\n
  &\quad\ -V'(x;\bm{\lambda})-V^{\prime*}(x;\bm{\lambda}).
  \label{H'}
\end{align}
This in turn implies that the twisted eigenfunction
$\tilde{\phi}_{\text{v}}(x;\bm{\lambda})$
\begin{equation}
  \tilde{\phi}_{\text{v}}(x;\bm{\lambda})\eqdef
  \phi_{\text{v}}\bigl(x;\mathfrak{t}(\bm{\lambda})\bigr)
  \quad(\text{v}\in\mathbb{Z}_{\ge0}),
\end{equation}
satisfies the original Schr\"odinger equation with
$\tilde{\mathcal{E}}_{\text{v}}(\bm{\lambda})$:
\begin{align}
  \mathcal{H}'(\bm{\lambda})\tilde{\phi}_{\text{v}}(x;\bm{\lambda})
  &=\mathcal{E}'_{\text{v}}(\bm{\lambda})
  \tilde{\phi}_{\text{v}}(x;\bm{\lambda}),\quad
  \mathcal{E}'_{\text{v}}(\bm{\lambda})\eqdef
  \mathcal{E}_{\text{v}}\bigl(\mathfrak{t}(\bm{\lambda})\bigr)\\
  &\Downarrow\n
  \mathcal{H}(\bm{\lambda})\tilde{\phi}_{\text{v}}(x;\bm{\lambda})
  &=\tilde{\mathcal{E}}_{\text{v}}(\bm{\lambda})
  \tilde{\phi}_{\text{v}}(x;\bm{\lambda}),\quad
  \tilde{\mathcal{E}}_{\text{v}}(\bm{\lambda})
  \eqdef\alpha(\bm{\lambda})
  \mathcal{E}_{\text{v}}\bigl(\mathfrak{t}(\bm{\lambda})\bigr)
  +\alpha'(\bm{\lambda}).
  \label{Etv}
\end{align}
If the following condition
\begin{equation}
  \tilde{\mathcal{E}}_{\text{v}}(\bm{\lambda})
  =\mathcal{E}_{-\text{v}-1}(\bm{\lambda})
  \label{vminv}
\end{equation}
is satisfied, the twisted eigenfunction
$\tilde{\phi}_{\text{v}}(x;\bm{\lambda})$ is called a {\em pseudo virtual
state wave function\/}.

For the Wilson and the Askey-Wilson polynomials, the appropriate twisting is:
\begin{align}
  \mathfrak{t}(\bm{\lambda})
  &\eqdef(1-\lambda_1,1-\lambda_2,1-\lambda_3,1-\lambda_4),\n
  &\Bigl(\ \text{or}\quad\left\{
  \begin{array}{ll}
  a_j\to 1-a_j&:\text{W}\\
  a_j\to qa_j^{-1}&:\text{AW}
  \end{array}\right.
  \ \ (j=1,\ldots,4)\ \Bigr),
  \label{WAWtwist}
\end{align}
with
\begin{align} 
  &\alpha(\bm{\lambda})=\left\{
  \begin{array}{ll}
  1&:\text{W}\\
  b_4q^{-2}&:\text{AW}
  \end{array}\right.,\quad
  \alpha'(\bm{\lambda})=\mathcal{E}_{-1}(\bm{\lambda})=\left\{
  \begin{array}{ll}
  -(b_1-2)&:\text{W}\\[2pt]
  -(1-q)(1-b_4q^{-2})&:\text{AW}
  \end{array}\right.,\\
  &\tilde{\mathcal{E}}_{\text{v}}(\bm{\lambda})
  =\mathcal{E}_{-\text{v}-1}(\bm{\lambda})
  =\left\{
  \begin{array}{ll}
  -(\text{v}+1)(b_1-\text{v}-2)&:\text{W}\\[2pt]
  -(1-q^{\text{v}+1})(1-b_4q^{-\text{v}-2})&:\text{AW}
  \end{array}\right..
\end{align}
The pseudo virtual state wave function $\tilde{\phi}_{\text{v}}$ reads
\begin{align}
  \tilde{\phi}_{\text{v}}(x;\bm{\lambda})
  &=\tilde{\phi}_0(x;\bm{\lambda})\check{\xi}_ {\text{v}}(x;\bm{\lambda}),
  \label{psvfac}\\
  \tilde{\phi}_0(x;\bm{\lambda})
  &\eqdef{\phi}_0\bigl(x;\mathfrak{t}(\bm{\lambda})\bigr),\quad
  \check{\xi}_{\text{v}}(x;\bm{\lambda})\eqdef
  \xi_{\text{v}}\bigl(\eta(x);\bm{\lambda}\bigr)\eqdef
  \check{P}_{\text{v}}\bigl(x;\mathfrak{t}(\bm{\lambda})\bigr)
  =P_{\text{v}}\bigl(\eta(x);\mathfrak{t}(\bm{\lambda})\bigr).
  \label{psvfac2}
\end{align}
The twisted potential is linearly related to the original potential by
\begin{equation}
  V'(x;\bm{\lambda})
  =\alpha(\bm{\lambda})^{-1}\frac{\varphi(x-i\gamma)}{\varphi(x)}
  V^*(x-i\gamma;\bm{\lambda}),
  \label{V'Vs}
\end{equation}
in which the auxiliary function $\varphi(x)$ is defined in \eqref{etadef}.

%%%%%%%%%%%%%%%%%%%%%%%%%%%%%%%%%%%%%%%%%%%%%%%%%%%%%%%%%%%%%%%
%                                                             %
%  3. Casoratian Identities for the Equivalence between       %
%     States Adding and Deleting Transformations              %
%                                                             %
%%%%%%%%%%%%%%%%%%%%%%%%%%%%%%%%%%%%%%%%%%%%%%%%%%%%%%%%%%%%%%%
\section{Casoratian Identities for the Equivalence between\\
Eigenstates Adding and Deleting Transformations}
\label{sec:Cas}

The main tool for deriving these identities is multiple Darboux
(Darboux-Crum) transformations, in terms of which various deformations of
solvable quantum mechanics are obtained.
In discrete quantum mechanics \cite{os15,gos}, as demonstrated for the
multi-indexed Wilson and Askey-Wilson polynomial cases \cite{os27},
the deformed potential functions and the deformed eigenfunctions etc can
be expressed neatly  by the Casoratians, which are the discrete analogues
of the Wronskians.

%%%%%%%%%%%%%%%%%%%%%%%%%%%%%%%%%%%%%%%%%%%%%%
%                                            %
% 3.1. Casoratian formulas                   %
%                                            %
%%%%%%%%%%%%%%%%%%%%%%%%%%%%%%%%%%%%%%%%%%%%%%
\subsection{Casoratian formulas}
\label{sec:casform}

First let us summarise the definitions and various properties of Casoratians.
The Casorati determinant of a set of $n$ functions $\{f_j(x)\}$ is defined by
\begin{equation}
  \text{W}_{\gamma}[f_1,\ldots,f_n](x)
  \eqdef i^{\frac12n(n-1)}
  \det\Bigl(f_k\bigl(x^{(n)}_j\bigr)\Bigr)_{1\leq j,k\leq n},\quad
  x_j^{(n)}\eqdef x+i(\tfrac{n+1}{2}-j)\gamma,
  \label{Wdef}
\end{equation}
(for $n=0$, we set $\text{W}_{\gamma}[\cdot](x)=1$),
which satisfies identities
\begin{align}
  &\text{W}_{\gamma}[f_1,\ldots,f_n]^*(x)
  =\text{W}_{\gamma}[f_1^*,\ldots,f_n^*](x),\\
  &\text{W}_{\gamma}[gf_1,gf_2,\ldots,gf_n]
  =\prod_{j=1}^ng\bigl(x^{(n)}_j\bigr)\cdot
  \text{W}_{\gamma}[f_1,f_2,\ldots,f_n](x),
  \label{dWformula1}\\
  &\text{W}_{\gamma}\bigl[\text{W}_{\gamma}[f_1,f_2,\ldots,f_n,g],
  \text{W}_{\gamma}[f_1,f_2,\ldots,f_n,h]\,\bigr](x)\n
  &=\text{W}_{\gamma}[f_1,f_2,\ldots,f_n](x)\,
  \text{W}_{\gamma}[f_1,f_2,\ldots,f_n,g,h](x)
  \quad(n\geq 0).
  \label{dWformula2}
\end{align}

%%%%%%%%%%%%%%%%%%%%%%%%%%%%%%%%%%%%%%%%%%%%%%%%%%%
%                                                 %
% 3.2. Eigenstates adding Darboux transformations %
%                                                 %
%%%%%%%%%%%%%%%%%%%%%%%%%%%%%%%%%%%%%%%%%%%%%%%%%%%
\subsection{Eigenstates adding Darboux transformations}
\label{sec:adddarb}

Now let us consider the deformation of the original system
\eqref{H}--\eqref{ortrel} by multiple Darboux transformations in terms of
$M$ pseudo virtual state wave functions indexed by the degrees of their
polynomial part wave functions.
Let $\mathcal{D}\eqdef\{d_1,d_2,\ldots,d_M\}$ ($d_j\in\mathbb{Z}_{\ge0}$)
be a set of distinct non-negative integers and we use the pseudo virtual
state wave functions $\{\tilde{\phi}_{d_j}(x;\bm{\lambda})\}$, $j=1,\ldots,M$
in this order.
In the formulas below \eqref{Hd1..ds}--\eqref{HPhi},
\eqref{hamhatA}--\eqref{hatAphi}, the parameter ($\bm{\lambda}$) dependence
is suppressed for simplicity of presentation.
The algebraic structure of the multiple Darboux transformations is the same
when the virtual or pseudo virtual state wave functions or the actual
eigenfunctions are used as {\em seed solutions\/}.
The system obtained after $s$ steps of Darboux transformations in terms of
pseudo virtual state wave functions labeled by $\{d_1,\ldots,d_s\}$
($s\geq 1$), is
\begin{align}
  &\mathcal{H}_{d_1\ldots d_s}\eqdef
  \hat{\mathcal{A}}_{d_1\ldots d_s}\hat{\mathcal{A}}_{d_1\ldots d_s}^{\dagger}
  +\tilde{\mathcal{E}}_{d_s},
  \label{Hd1..ds}\\
  &\hat{\mathcal{A}}_{d_1\ldots d_s}\eqdef
  i\bigl(e^{\frac{\gamma}{2}p}\sqrt{\hat{V}_{d_1\ldots d_s}^*(x)}
  -e^{-\frac{\gamma}{2}p}\sqrt{\hat{V}_{d_1\ldots d_s}(x)}\,\bigr),\n
  &\hat{\mathcal{A}}_{d_1\ldots d_s}^{\dagger}\eqdef
  -i\bigl(\sqrt{\hat{V}_{d_1\ldots d_s}(x)}\,e^{\frac{\gamma}{2}p}
  -\sqrt{\hat{V}_{d_1\ldots d_s}^*(x)}\,e^{-\frac{\gamma}{2}p}\bigr),\\
  &\hat{V}_{d_1\ldots d_s}(x)\eqdef
  \sqrt{V(x-i\tfrac{s-1}{2}\gamma)V^*(x-i\tfrac{s+1}{2}\gamma)}\n
  &\phantom{\hat{V}_{d_1\ldots d_s}(x)\eqdef}\times
  \frac{\text{W}_{\gamma}[\tilde{\phi}_{d_1},\ldots,\tilde{\phi}_{d_{s-1}}]
  (x+i\frac{\gamma}{2})}
  {\text{W}_{\gamma}[\tilde{\phi}_{d_1},\ldots,\tilde{\phi}_{d_{s-1}}]
  (x-i\frac{\gamma}{2})}\,
  \frac{\text{W}_{\gamma}[\tilde{\phi}_{d_1},\ldots,\tilde{\phi}_{d_s}]
  (x-i\gamma)}
  {\text{W}_{\gamma}[\tilde{\phi}_{d_1},\ldots,\tilde{\phi}_{d_s}](x)},
  \label{Vhd1..ds}\\
  &\phi_{d_1\ldots d_s\,n}(x)\eqdef
  \hat{\mathcal{A}}_{d_1\ldots d_s}\phi_{d_1\ldots d_{s-1}\,n}(x)
  \ \ (n=0,1,2,\ldots),\n
  &\tilde{\phi}_{d_1\ldots d_s\,\text{v}}(x)\eqdef
  \hat{\mathcal{A}}_{d_1\ldots d_s}
  \tilde{\phi}_{d_1\ldots d_{s-1}\,\text{v}}(x)
  \ (\text{v}\in\mathcal{D}\backslash\{d_1,\ldots,d_s\}),\\
  &\mathcal{H}_{d_1\ldots d_s}\phi_{d_1\ldots d_s\,n}(x)
  =\mathcal{E}_n\phi_{d_1\ldots d_s\,n}(x)
  \ \ (n=0,1,2,\ldots),\n
  &\mathcal{H}_{d_1\ldots d_s}\tilde{\phi}_{d_1\ldots d_s\,\text{v}}(x)
  =\tilde{\mathcal{E}}_\text{v}\tilde{\phi}_{d_1\ldots d_s\,\text{v}}(x)
  \ \ (\text{v}\in\mathcal{D}\backslash\{d_1,\ldots,d_s\}).
  \label{Hd1..dsphid1..ds=}
\end{align}
The eigenfunctions and the pseudo virtual state wave functions in all steps
are `real' by construction,
$\phi_{d_1\ldots d_s\,n}^*(x)=\phi_{d_1\ldots d_s\,n}(x)$,
$\tilde{\phi}_{d_1\ldots d_s\,\text{v}}^*(x)
=\tilde{\phi}_{d_1\ldots d_s\,\text{v}}(x)$
and they have Casoratian expressions:
\begin{align}
  &\phi_{d_1\ldots d_s\,n}(x)=A(x)
  \text{W}_{\gamma}[\tilde{\phi}_{d_1},\ldots,\tilde{\phi}_{d_s},\phi_n](x),
  \n
  &\tilde{\phi}_{d_1\ldots d_s\,\text{v}}(x)=A(x)
  \text{W}_{\gamma}[\tilde{\phi}_{d_1},\ldots,\tilde{\phi}_{d_s},
  \tilde{\phi}_{\text{v}}](x),
  \label{phid1..dsn}\\
  &\quad A(x)=\left(
  \frac{\sqrt{\prod_{j=0}^{s-1}V(x+i(\frac{s}{2}-j)\gamma)
  V^*(x-i(\frac{s}{2}-j)\gamma)}}
  {\text{W}_{\gamma}[\tilde{\phi}_{d_1},\ldots,\tilde{\phi}_{d_s}]
  (x-i\frac{\gamma}{2})
  \text{W}_{\gamma}[\tilde{\phi}_{d_1},\ldots,\tilde{\phi}_{d_s}]
  (x+i\frac{\gamma}{2})}\right)^{\frac12}.
  \nonumber
\end{align}
These are essentially the same as those obtained for the multi-indexed
polynomials as given in (2.18)--(2.24) of \cite{os27}, which have been
derived in terms of the virtual state wave functions.

One marked difference from the multi-indexed polynomials case, in which
virtual state wave functions are used, is the appearance of {\em new
eigenstates\/} below the original ground state ($\tilde{\mathcal E}_{d_j}<0$)
as many as those used pseudo virtual state wave functions:
\begin{align}
  &\breve{\Phi}_{d_1\ldots d_s;d_j}(x)
  \eqdef\mathcal{C}_s(x)\times
  \Bigl(\prod_{k=0}^{s-1}V\bigl(x+i(\tfrac{s}{2}-k)\gamma\bigr)
  V^*\bigl(x-i(\tfrac{s}{2}-k)\gamma\bigr)\Bigr)^{-\frac14}\n
  &\phantom{\breve{\Phi}_{d_1\ldots d_s;d_j}(x)\eqdef}\times
  \frac{\text{W}_{\gamma}[\tilde{\phi}_{d_1},\ldots,\breve{\tilde{\phi}}_{d_j},
  \ldots,\tilde{\phi}_{d_s}](x)}
  {\sqrt{\text{W}_{\gamma}[\tilde{\phi}_{d_1},\ldots,\tilde{\phi}_{d_s}]
  (x-i\frac{\gamma}{2})
  \text{W}_{\gamma}[\tilde{\phi}_{d_1},\ldots,\tilde{\phi}_{d_s}]
  (x+i\frac{\gamma}{2})}},
  \label{Phi}\\
  &\mathcal{H}_{d_1\ldots d_s}\breve{\Phi}_{d_1\ldots d_s;d_j}(x)
  =\tilde{\mathcal{E}}_{d_j}\breve{\Phi}_{d_1\ldots d_s;d_j}(x)
  \quad(j=1,2,\ldots,s),
  \label{HPhi}
\end{align}
in which $\mathcal{C}_s(x)$ is given by
\begin{equation}
  \mathcal{C}_s(x)=
  \frac{\phi_0(x;\bm{\lambda}-s\bm{\delta})
  \phi_0(x;\mathfrak{t}(\bm{\lambda}-s\bm{\delta}))}
  {\varphi(x)},
  \label{Csform}
\end{equation}
satisfying the pseudo constant condition
$\mathcal{C}_s(x-i\gamma)=\mathcal{C}_s(x)$.
In the numerator of \eqref{Phi},
$\text{W}_{\gamma}[\tilde{\phi}_{d_1},\ldots,\breve{\tilde{\phi}}_{d_j},
\ldots,\tilde{\phi}_{d_s}](x)$ means that $\tilde{\phi}_{d_j}$ is
excluded from the Casoratian.
Since the Hamiltonian $\mathcal{H}_{d_1\ldots d_s}$ can be rewritten as
\begin{equation}
  \mathcal{H}_{d_1\ldots d_s}
  =\hat{\mathcal{A}}_{d_1\ldots\breve{d_j}\ldots d_sd_j}
  \hat{\mathcal{A}}_{d_1\ldots\breve{d_j}\ldots d_sd_j}^{\dagger}
  +\tilde{\mathcal{E}}_{d_j},
  \label{hamhatA}
\end{equation}
the new eigenstates are the zero modes of the operator
$\hat{\mathcal{A}}_{d_1\ldots\breve{d_j}\ldots d_sd_j}^{\dagger}$:
\begin{equation}
  \hat{\mathcal{A}}_{d_1\ldots\breve{d_j}\ldots d_sd_j}^{\dagger}
  \breve{\Phi}_{d_1\ldots d_s;d_j}(x)=0\quad(j=1,2,\ldots,s).
  \label{hatAphi}
\end{equation}

For the elementary Darboux transformation, $s=1$, the above zero mode
\eqref{Phi} reads simply
\begin{equation}
  \breve{\Phi}_{d_1;d_1}(x)
  \propto
  \frac{\phi_0(x;\bm{\lambda}-\bm{\delta})}
  {\sqrt{\check{\xi}_{d_1}(x-i\tfrac{\gamma}2;\bm{\lambda})
  \check{\xi}_{d_1}(x+i\tfrac{\gamma}2;\bm{\lambda})}},
  \label{onezero}
\end{equation}
for which the discrete symmetry relation \eqref{VV'rel}, the zero mode
equation \eqref{zero} and the shape covariant relation of $\phi_0$
\eqref{phi0shape} are used.
It is straightforward to verify
$\hat{\mathcal{A}}_{d_1}^{\dagger}\breve{\Phi}_{d_1;d_1}(x)=0$.
This wave function indeed describes an eigenstate of $\mathcal{H}_{d_1}$,
so long as the polynomial $\check{\xi}_ {d_1}(x;\bm{\lambda})$ does not
have zeros in a certain domain (see \S\,3.4 of \cite{os27}, Appendix A of
\cite{os13}) and the parameter ranges are narrowed than the original theory.
For example, for the Wilson and Askey-Wilson, they are
\begin{equation}
  d_1:\ \text{even};\qquad
  \text{W}:\ \text{Re}\,a_j>\tfrac12,\quad
  \text{AW}:\ |a_j|<q^{\frac12}\quad(j=1,\ldots,4),
  \label{onenarr}
\end{equation}
in contrast with the original parameter range given in \eqref{rangeorg}.

It is illuminating to compare the above zero mode \eqref{onezero} with the
corresponding ones in the ordinary quantum mechanics. For example, for the
P\"oschl-Teller potential $\bigl(\bm{\lambda}=(g,h)$, $\bm{\delta}=(1,1)\bigr)$,
\begin{equation*}
  U(x;\bm{\lambda})=\frac{g(g-1)}{\sin^2x}+\frac{h(h-1)}{\cos^2x}-(g+h)^2,
\end{equation*}
the pseudo virtual state wave function and the corresponding zero mode,
which is simply a reciprocal, are
$\bigl(\mathfrak{t}(\bm{\lambda})=(1-g,1-h)\bigr)$ \cite{os29}:
\begin{align*} 
  \tilde{\phi}_\text{v}(x;\bm{\lambda})
  &=(\sin x)^{1-g}(\cos x)^{1-h}
  P_\text{v}^{(\frac12-g,\frac12-h)}(\cos2x),\\[4pt]
  \tilde{\phi}_\text{v}(x;\bm{\lambda})^{-1}
  &=\frac{(\sin x)^{g-1}(\cos x)^{h-1}}
  {P_\text{v}^{(\frac12-g,\frac12-h)}(\cos2x)}
  =\frac{\phi_0(x;\bm{\lambda}-\bm{\delta})}
  {P_\text{v}^{(\frac12-g,\frac12-h)}(\cos2x)}.
\end{align*}

It should be stressed that for the virtual state wave functions \cite{os27},
the function $\mathcal{C}_s(x)$ \eqref{Csform} is {\em not a pseudo
constant\/} $\mathcal{C}_s(x)\neq\mathcal{C}_s(x-i\gamma)$. That is,
in the Darboux transformations in terms of virtual states, the wave function
\eqref{Phi} with \eqref{Csform} does not satisfy the Schr\"odinger equation
\eqref{HPhi}.
The function $\mathcal{C}_s(x)$ \eqref{Csform} plays an important role to
guarantee for the newly added eigenstates \eqref{Phi} to belong to the
proper Hilbert space of the deformed Hamiltonian
$\mathcal{H}_{d_1\ldots d_s}$ \eqref{Hd1..ds}.

Let us introduce appropriate notation for the quantities after the full
deformation using the $M$ pseudo virtual state wave functions specified by
$\mathcal{D}\eqdef\{d_1,d_2,\ldots,d_M\}$ ($d_j\in\mathbb{Z}_{\ge0}$).
We use simplified notation
$\mathcal{H}_{d_1\ldots d_M}=\mathcal{H}_{\mathcal{D}}$,
$\hat{\mathcal{A}}_{d_1\ldots d_M}=\hat{\mathcal{A}}_{\mathcal{D}}$,
$\hat{V}_{d_1\ldots d_M}(x)=\hat{V}_{\mathcal{D}}(x)$,
$\phi_{d_1\ldots d_M\,n}(x)=\phi_{\mathcal{D}\,n}(x)$,
$\breve{\Phi}_{d_1\ldots d_M;d_j}(x)=\breve{\Phi}_{\mathcal{D};d_j}(x)$ etc,
\begin{equation}
  \mathcal{H}_{\mathcal{D}}
  =\hat{\mathcal{A}}_{\mathcal{D}}\hat{\mathcal{A}}_{\mathcal{D}}^\dagger
  +\tilde{\mathcal E}_{d_M}.
  \label{HDdef}
\end{equation}
The Casoratians of eigenfunctions, the pseudo virtual state wave functions
and mixed ones are factorised into a polynomial in $\eta(x)$ (the sinusoidal
coordinate) and a kinematical factor.
For eigenfunctions only we have
\begin{align}
  &\text{W}_{\gamma}[\phi_{d_1},\phi_{d_2},\ldots,\phi_{d_M}](x;\bm{\lambda})
  =\bar{A}_{\mathcal{D}}(x;\bm{\lambda})
  \bar{\Xi}_{\mathcal{D}}\bigl(\eta(x);\bm{\lambda}\bigr),\\
  &\bar{A}_{\mathcal{D}}(x;\bm{\lambda})
  \eqdef\prod_{j=1}^M\phi_0(x^{(M)}_j;\bm{\lambda})\cdot\varphi_M(x),\\
  &\bar{\Xi}_{\mathcal{D}}\bigl(\eta(x);\bm{\lambda}\bigr)
  \eqdef\varphi_M(x)^{-1}\,
  \text{W}_{\gamma}[\check{P}_{d_1},\check{P}_{d_2},\ldots,\check{P}_{d_M}]
  (x;\bm{\lambda}).
\end{align}
Here we use the symbol $x^{(n)}_j=x+i(\frac{n+1}{2}-j)$ as introduced
in \eqref{Wdef} and the auxiliary function $\varphi_M(x)$ \cite{gos} is
defined by:
\begin{align}
  \varphi_M(x)&\eqdef
  \varphi(x)^{[\frac{M}{2}]}\prod_{k=1}^{M-2}
  \bigl(\varphi(x-i\tfrac{k}{2}\gamma)\varphi(x+i\tfrac{k}{2}\gamma)
  \bigr)^{[\frac{M-k}{2}]}\n
  &=\prod_{1\leq j<k\leq M}
  \frac{\eta(x^{(M)}_j)-\eta(x^{(M)}_k)}
  {\varphi(i\frac{j}{2}\gamma)}
  \times\left\{
  \begin{array}{ll}
  1&:\text{W}\\
  (-2)^{\frac12M(M-1)}&:\text{AW}
  \end{array}\right.,
  \label{varphiMdef}
\end{align}
and $\varphi_0(x)=\varphi_1(x)=1$.
Here $[x]$ denotes the greatest integer not exceeding $x$.\\
The Casoratian containing  the $M$ pseudo virtual state wave functions only
reads:
\begin{align}
  &\text{W}_{\gamma}[\tilde{\phi}_{d_1},\tilde{\phi}_{d_2},\ldots,
  \tilde{\phi}_{d_M}](x;\bm{\lambda})
  =A_{\mathcal{D}}(x;\bm{\lambda})
  \Xi_{\mathcal{D}}\bigl(\eta(x);\bm{\lambda}\bigr),\\
  &A_{\mathcal{D}}(x;\bm{\lambda})
  \eqdef\prod_{j=1}^M\tilde{\phi}_0\bigl(x^{(M)}_j;\bm{\lambda}\bigr)
  \cdot\varphi_M(x),\\
  &\Xi_{\mathcal{D}}\bigl(\eta(x);\bm{\lambda}\bigr)
  \eqdef\varphi_M(x)^{-1}\,
  \text{W}_{\gamma}[\check{\xi}_{d_1},\check{\xi}_{d_2},\ldots,
  \check{\xi}_{d_M}](x;\bm{\lambda}).
\end{align}
The Casoratian containing the $M$ pseudo virtual state wave functions and
one eigenfunction reads:
\begin{align}
  &\text{W}_{\gamma}[\tilde{\phi}_{d_1},\tilde{\phi}_{d_2},\ldots,
  \tilde{\phi}_{d_M},\phi_n](x;\bm{\lambda})
  =A_{\mathcal{D},n}(x;\bm{\lambda})
  P_{\mathcal{D},n}\bigl(\eta(x);\bm{\lambda}\bigr),\\
  &A_{\mathcal{D},n}(x;\bm{\lambda})
  \eqdef\prod_{j=1}^{M+1}\tilde{\phi}_0\bigl(x^{(M+1)}_j;\bm{\lambda}\bigr)
  \cdot\nu(x;\bm{\lambda}-M\bm{\delta})\varphi_{M+1}(x),\\
  &P_{\mathcal{D},n}\bigl(\eta(x);\bm{\lambda}\bigr)
  \eqdef\varphi_{M+1}(x)^{-1}i^{\frac12M(M+1)}\left|
  \begin{array}{llll}
  \vec{X}^{(M+1)}_{d_1}&\cdots&\vec{X}^{(M+1)}_{d_{M}}&
  \vec{Z}^{(M+1)}_n\\
  \end{array}\right|,
\end{align}
where
\begin{align}
  &\bigl(\vec{X}^{(M+1)}_{\text{v}}\bigr)_j\eqdef
  \check{\xi}_{\text{v}}\bigl(x^{(M+1)}_j;\bm{\lambda}\bigr)\quad
  (1\leq j\leq M+1),\\
  &\bigl(\vec{Z}^{(M+1)}_n\bigr)_j\eqdef
  r_j(x^{(M+1)}_j;\bm{\lambda},M+1)\check{P}_n(x^{(M+1)}_j;\bm{\lambda})\quad
  (1\leq j\leq M+1),\\
  &\nu(x;\bm{\lambda})\eqdef\frac{\phi_0(x;\bm{\lambda})}
  {\tilde{\phi}_0(x;\bm{\lambda})},\\
  &r_j(x^{(M+1)}_j;\bm{\lambda},M+1)\eqdef
  \frac{\nu(x^{(M+1)}_j;\bm{\lambda})}
  {\nu(x;\bm{\lambda}-M\bm{\delta})}\quad(1\leq j\leq M+1),\n
  &\qquad
  \propto\left\{
  \begin{array}{ll}
  \prod_{k=1}^4(a_k-\frac{M}{2}+ix)_{j-1}(a_k-\frac{M}{2}-ix)_{M+1-j}
  &:\text{W}\\[3pt]
  e^{2ix(M+2-2j)}\prod_{k=1}^4(a_kq^{-\frac{M}{2}}e^{ix};q)_{j-1}
  (a_kq^{-\frac{M}{2}}e^{-ix};q)_{M+1-j}
  &:\text{AW}
  \end{array}\right..
\end{align}
In these expressions $\bar{\Xi}_{\mathcal{D}}(\eta;\bm{\lambda})$,
$\Xi_{\mathcal{D}}(\eta;\bm{\lambda})$,
$P_{\mathcal{D},n}(\eta;\bm{\lambda})$  are polynomials in $\eta$ and their
degrees are generically $\ell_{\mathcal{D}}$, $\ell_{\mathcal{D}}$,
$\ell_{\mathcal{D}}+M+n$, respectively. Here $\ell_{\mathcal{D}}$ is defined by
\begin{equation}
  \ell_{\mathcal{D}}\eqdef\sum_{j=1}^Md_j-\frac12M(M-1).
  \label{ellD}
\end{equation}
The kinematical factors $\bar{A}_{\mathcal{D}}$, $A_{\mathcal{D}}$,
$\mathcal{A}_{\mathcal{D},n}$ depend on $M$ but they are independent of
the explicit choices of the degrees $\{d_j\}$.
There are obvious relations
\begin{equation}
  A_{\mathcal{D}}(x;\bm{\lambda})
  =\bar{A}_{\mathcal{D}}\bigl(x;\mathfrak{t}(\bm{\lambda})\bigr),\quad
  \Xi_{\mathcal{D}}(\eta;\bm{\lambda})
  =\bar{\Xi}_{\mathcal{D}}\bigl(\eta;\mathfrak{t}(\bm{\lambda})\bigr),
\end{equation}
reflecting the fact that the pseudo virtual state wave functions are defined by
twisting \eqref{psvfac2}.

The deformed eigenfunctions $\phi_{\mathcal{D}\,n}$, the newly added
eigenfunctions $\breve{\Phi}_{\mathcal{D};d_j}$ and the deformed potential
function $\hat{V}_{\mathcal{D}}$ are expressed neatly in terms of the above
quantities with $\mathcal{D}'$ defined by
$\mathcal{D}'\eqdef\{d_1,\ldots,d_{M-1}\}$:
\begin{align}
  \phi_{\mathcal{D}\,n}(x;\bm{\lambda})
  &\propto
  \psi_{\mathcal{D}}(x;\bm{\lambda})
  \check{P}_{\mathcal{D},n}(x;\bm{\lambda})\quad(n=0,1,\ldots),
  \label{eigXi1}\\
  \psi_{\mathcal{D}}(x;\bm{\lambda})
  &\eqdef\frac{\phi_0(x;\bm{\lambda}-M\bm{\delta})}
  {\sqrt{\check{\Xi}_{\mathcal{D}}(x-i\frac{\gamma}{2};\bm{\lambda})
  \check{\Xi}_{\mathcal{D}}(x+i\frac{\gamma}{2};\bm{\lambda})}},\\
  \breve{\Phi}_{\mathcal{D};d_j}(x;\bm{\lambda})
  &\propto
  \psi_{\mathcal{D}}(x;\bm{\lambda})\,
  \check{\Xi}_{d_1\ldots\breve{d_j}\ldots d_M}(x;\bm{\lambda})
  \quad(j=1,\ldots,M),
  \label{eigXi2}\\
  \hat{V}_{\mathcal{D}}(x;\bm{\lambda})
  &=\kappa^{-M}V^*(x-i\tfrac{\gamma}{2};\bm{\lambda}-M\bm{\delta})
  \frac{\check{\Xi}_{\mathcal{D}'}(x+i\frac{\gamma}{2};\bm{\lambda})}
  {\check{\Xi}_{\mathcal{D}'}(x-i\frac{\gamma}{2};\bm{\lambda})}
  \frac{\check{\Xi}_{\mathcal{D}}(x-i\gamma;\bm{\lambda})}
  {\check{\Xi}_{\mathcal{D}}(x;\bm{\lambda})}.
  \label{VXi}
\end{align}
Here, as before, we have used the notation
$\check{\Xi}_{\mathcal{D}}(x;\bm{\lambda})
=\Xi_{\mathcal{D}}(\eta(x);\bm{\lambda})$, etc.

The shape invariance of the original theory implies relations 
\begin{equation}
  \bar{\Xi}_{\{d_1,\ldots,d_M,0\}}(\eta;\bm{\lambda})
  \propto\bar{\Xi}_{\{d_1-1,\ldots,d_M-1\}}(\eta;\bm{\lambda}+\bm{\delta}),
  \label{PoD0=}
\end{equation}
which are the difference analogues of the relations (4.33) of \cite{os29}.
They are derived based on the forward shift relation \eqref{FP=,BP=},
the property of the Casoratian
\begin{equation}
  \text{W}_{\gamma}[1,f_1,\ldots,f_n](x)
  =\text{W}_{\gamma}[F_1,\ldots,F_n](x),\quad
  F_j(x)\eqdef-i\bigl(f_j(x+i\tfrac{\gamma}{2})
  -f_j(x-i\tfrac{\gamma}{2})\bigr),
\end{equation}
and the property of $\varphi_M(x)$,
\begin{equation}
  \varphi_{M+1}(x)=\varphi_M(x)\prod_{j=1}^M\varphi(x^{(M)}_j)
  \quad(M\geq 0).
\end{equation}
By repeating \eqref{PoD0=}, one arrives at
\begin{equation}
  \bar{\Xi}_{\{0,1,\ldots,n\}}(\eta;\bm{\lambda})
  \propto\bar{\Xi}_{\{0,1,\ldots,n-1\}}(\eta;\bm{\lambda}+\bm{\delta})
  \propto\cdots\propto\bar{\Xi}_{\{0\}}(\eta;\bm{\lambda}+n\bm{\delta})
  =\text{constant}.
  \label{Xi01..n}
\end{equation}

%%%%%%%%%%%%%%%%%%%%%%%%%%%%%%%%%%%%%%%%%%%%%%%%%%%%%
%                                                   %
% 3.3. Eigenstates deleting Darboux transformations %
%                                                   %
%%%%%%%%%%%%%%%%%%%%%%%%%%%%%%%%%%%%%%%%%%%%%%%%%%%%%
\subsection{Eigenstates deleting Darboux transformations}
\label{sec:deldarb}

In a previous publication \cite{os29} we have shown for various solvable
potentials in ordinary quantum mechanics that the eigenstates adding
Darboux transformations are dual to eigenstates deleting Krein-Adler
transformations with shifted parameters. The situation is the same for
various solvable theories in discrete quantum mechanics.
The correspondence among the added eigenstates specified by $\mathcal{D}$
and the deleted eigenstates $\bar{\mathcal D}$ with shifted parameter
$\bar{\bm \lambda}$ \eqref{barD} is depicted in Fig.\,1.

Let us introduce an integer $N$ and fix it to be not less than the maximum of
$\mathcal{D}$:
\begin{equation}
  N\ge\text{max}(\mathcal{D}).
\end{equation}
This determines a set of distinct non-negative integers
$\bar{\mathcal{D}}=\{0,1,\ldots,N\}\backslash
\{\bar{d}_1,\bar{d}_2,\ldots,\bar{d}_M\}$
together with the shifted parameters $\bar{\bm{\lambda}}$:
\begin{align}
  &\bar{\mathcal{D}}\eqdef\{0,1,\ldots,\breve{\bar{d}}_1,\ldots,
  \breve{\bar{d}}_2,\ldots,\breve{\bar{d}}_M,\ldots,N\}
  =\{e_1,e_2,\ldots,e_{N+1-M}\},\n
  &\bar{d}_j\eqdef N-d_j,\quad
  \bar{\bm{\lambda}}\eqdef\bm{\lambda}-(N+1)\bm{\delta}.
  \label{barD}
\end{align}
The eigenvalue $\mathcal{E}_n$ as a function of the parameters $\bm{\lambda}$
in general satisfies the relations:
\begin{align}
  \mathcal{E}_n(\bm{\lambda})-\mathcal{E}_{-N-1}(\bm{\lambda})
  &=\kappa^{-N-1}\mathcal{E}_{N+1+n}\bigl(\bar{\bm{\lambda}}\bigr),
  \label{enformula1}\\
  \mathcal{E}_{-\text{v}-1}(\bm{\lambda})-\mathcal{E}_{-N-1}(\bm{\lambda})
  &=\kappa^{-N-1}\mathcal{E}_{N-\text{v}}
  \bigl(\bar{\bm{\lambda}}\bigr).
  \label{enformula2}
\end{align}
The first relation \eqref{enformula1} says that $n$-th eigen level of
the original system corresponds to $(N+1+n)$-th level of the parameter
shifted system.
The second formula \eqref{enformula2} means that the state created by a
pseudo virtual state wave function $\tilde{\phi}_\text{v}$ is related to
$\bar{\text v}$-th level of the parameter shifted system.
These relations are the base of the duality depicted in Fig.\,1.
Among the newly created eigenfunctions the  lowest energy level $\mu$ is
given by
\begin{equation}
  \mu=\min\bigl(\mathbb{Z}_{\geq 0}\backslash\bar{\mathcal{D}}\bigr)
  =\min\{\bar{d}_1,\ldots,\bar{d}_M\}.
\end{equation}
The choice of the integer $N$ is not unique and the systems with different
$N$ are related by shape invariance.

\begin{figure}[htbp]
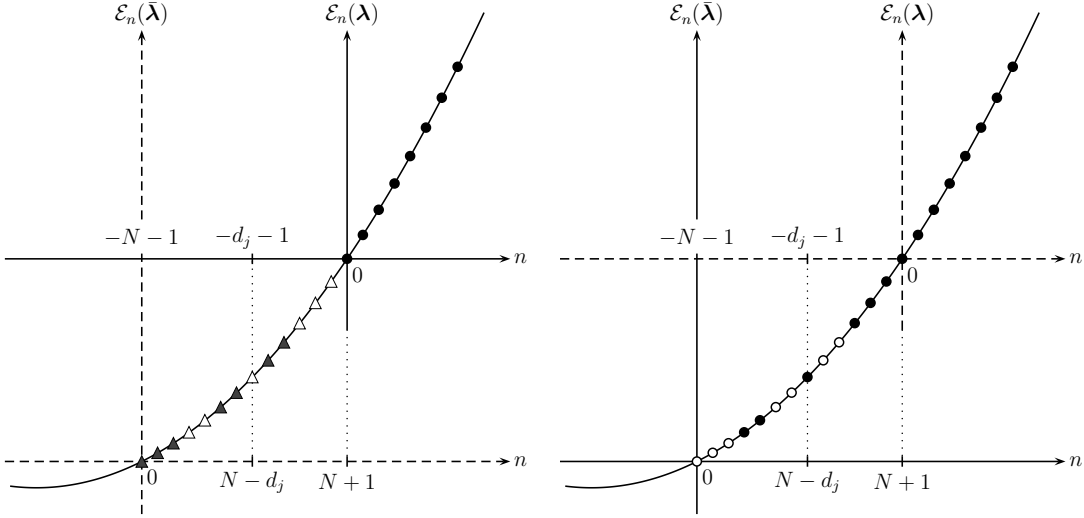

\begin{center}
  \scalebox{0.7}{\includegraphics{enfigIII_pvs.epsi}}\quad
  \scalebox{0.7}{\includegraphics{enfigIII_es.epsi}}
  \caption{The left represents the Darboux-Crum transformations in terms of
pseudo virtual states. The right corresponds to the Krein-Adler
transformations in terms of eigenstates with shifted parameters.
The black circles denote eigenstates.
The white circles in the right graphic denote deleted eigenstates.
The white triangles in the left graphic denote the pseudo virtual states
used in the Darboux-Crum transformations in \S\,\ref{sec:adddarb}.
The black triangles denote the unused pseudo virtual states.
}
\end{center}
\end{figure}

Let us denote the above eigenstate deleted system by
$\mathcal{H}^{\text{KA}}_{\bar{\mathcal{D}}}$,
$\mathcal{A}^{\text{KA}}_{\bar{\mathcal{D}}}$,
$V^{\text{KA}}_{\bar{\mathcal{D}}}(x)$, etc.
The general formulas of the Krein-Adler transformations \cite{os15,gos} provide:
\begin{align}
  \!\!\!&\mathcal{H}^{\text{KA}}_{\bar{\mathcal{D}}}
  =\mathcal{A}_{\bar{\mathcal{D}}}^{\text{KA}\,\dagger}
  \mathcal{A}^{\text{KA}}_{\bar{\mathcal{D}}}
  +\mathcal{E}_{\mu}(\bar{\bm{\lambda}}),\quad
  \mathcal{A}^{\text{KA}}_{\bar{\mathcal{D}}}=
  i\bigl(e^{\frac{\gamma}{2}p}\sqrt{V_{\bar{\mathcal{D}}}^{\text{KA}\,*}(x)}
  -e^{-\frac{\gamma}{2}p}\sqrt{V^{\text{KA}}_{\bar{\mathcal{D}}}(x)}\,\bigr),\n
  \!\!\!&\phantom{\mathcal{H}^{\text{KA}}_{\bar{\mathcal{D}}}
  =\mathcal{A}_{\bar{\mathcal{D}}}^{\text{KA}\,\dagger}
  \mathcal{A}^{\text{KA}}_{\bar{\mathcal{D}}}
  +\mathcal{E}_{\mu}(\bar{\bm{\lambda}}),\quad}
  \mathcal{A}^{\text{KA}\,\dagger}_{\bar{\mathcal{D}}}=
  -i\bigl(\sqrt{V^{\text{KA}}_{\bar{\mathcal{D}}}(x)}\,e^{\frac{\gamma}{2}p}
  -\sqrt{V_{\bar{\mathcal{D}}}^{\text{KA}\,*}(x)}\,
  e^{-\frac{\gamma}{2}p}\bigr),\\
  \!\!\!&V^{\text{KA}}_{\bar{\mathcal{D}}}(x)=
  \sqrt{V(x-i\tfrac{N+1-M}{2}\gamma;\bar{\bm{\lambda}})
  V^*(x-i\tfrac{N+3-M}{2}\gamma;\bar{\bm{\lambda}})}\n
  \!\!\!&\phantom{V^{\text{KA}}(x)=}\times
  \frac{\text{W}_{\gamma}[\phi_{e_1},\ldots,\phi_{e_{N+1-M}}]
  (x+i\frac{\gamma}{2};\bar{\bm{\lambda}})}
  {\text{W}_{\gamma}[\phi_{e_1},\ldots,\phi_{e_{N+1-M}}]
  (x-i\frac{\gamma}{2};\bar{\bm{\lambda}})}\,
  \frac{\text{W}_{\gamma}[\phi_{e_1},\ldots,\phi_{e_{N+1-M}},\phi_{\mu}]
  (x-i\gamma;\bar{\bm{\lambda}})}
  {\text{W}_{\gamma}[\phi_{e_1},\ldots,\phi_{e_{N+1-M}},\phi_{\mu}]
  (x;\bar{\bm{\lambda}})},\\
  \!\!\!&\Phi^{\text{KA}}_{\bar{\mathcal{D}}\,n}(x)
  =A^{\text{KA}}_{\bar{\mathcal{D}}}(x)
  \text{W}_{\gamma}[\phi_0,\phi_1,\ldots,\breve{\phi}_{\bar{d}_1},\ldots,
  \breve{\phi}_{\bar{d}_M},\ldots,\phi_N,\phi_{N+1+n}](x;\bar{\bm{\lambda}})
  \ \,(n=0,1,\ldots),\\
  \!\!\!&\breve{\Phi}^{\text{KA}}_{\bar{\mathcal{D}};d_j}(x)
  =A^{\text{KA}}_{\bar{\mathcal{D}}}(x)
  \text{W}_{\gamma}[\phi_0,\phi_1,\ldots,\breve{\phi}_{\bar{d}_1},\ldots,
  \phi_{\bar{d}_j},\ldots,\breve{\phi}_{\bar{d}_M},\ldots,\phi_N]
  (x;\bar{\bm{\lambda}})\ (j=1,\ldots,M),\!\!\!\\
  \!\!\!&\quad A^{\text{KA}}_{\bar{\mathcal{D}}}(x)=\left(
  \frac{\sqrt{\prod_{j=0}^{N-M}V(x+i(\frac{s}{2}-j)\gamma;\bar{\bm{\lambda}})
  V^*(x-i(\frac{s}{2}-j)\gamma;\bar{\bm{\lambda}})}}
  {\text{W}_{\gamma}[\phi_{e_1},\ldots,\phi_{e_{N+1-M}}]
  (x-i\frac{\gamma}{2};\bar{\bm{\lambda}})
  \text{W}_{\gamma}[\phi_{e_1},\ldots,\phi_{e_{N+1-M}}]
  (x+i\frac{\gamma}{2};\bar{\bm{\lambda}})}\right)^{\frac12}.
\end{align}
In terms of the polynomial $\bar{\Xi}_{\bar{\mathcal D}}$ the eigenfunctions
are expressed in a similar way as \eqref{eigXi1}--\eqref{eigXi2}:
\begin{align}
  &\Phi^{\text{KA}}_{\bar{\mathcal{D}}\,n}(x)=\kappa^{\frac14M(M-1)}
  \frac{\phi_0(x;\bm{\lambda}-M\bm{\delta})}{\sqrt{
  \check{\bar{\Xi}}_{\bar{\mathcal{D}}}(x-i\frac{\gamma}{2};\bar{\bm{\lambda}})
  \check{\bar{\Xi}}_{\bar{\mathcal{D}}}(x+i\frac{\gamma}{2};\bar{\bm{\lambda}})
  }}\,
  \check{\bar{\Xi}}_{\bar{\mathcal{D}}\,N+1+n}(x;\bar{\bm{\lambda}}),
  \label{xiDn1}\\
  &\breve{\Phi}^{\text{KA}}_{\bar{\mathcal{D}};d_j}(x)=\kappa^{\frac14M(M-1)}
  \frac{\phi_0(x;\bm{\lambda}-M\bm{\delta})}{\sqrt{
  \check{\bar{\Xi}}_{\bar{\mathcal{D}}}(x-i\frac{\gamma}{2};\bar{\bm{\lambda}})
  \check{\bar{\Xi}}_{\bar{\mathcal{D}}}(x+i\frac{\gamma}{2};\bar{\bm{\lambda}})
  }}\,
  \check{\bar{\Xi}}_{01\ldots\breve{\bar{d}}_1\ldots\bar{d}_j\ldots
  \breve{\bar{d}}_M\ldots N}(x;\bar{\bm{\lambda}}),
   \label{xiDn2}
\end{align}
in which
$\check{\bar{\Xi}}_{01\ldots\breve{\bar{d}}_1\ldots\bar{d}_j\ldots
\breve{\bar{d}}_M\ldots N}(x;\bar{\bm{\lambda}})
=\pm\check{\bar{\Xi}}_{\bar{\mathcal{D}}\,\bar{d}_j}(x;\bar{\bm{\lambda}})$.
Let us take, without loss of generality, $d_1<\cdots<d_M$.
This means that $\mu=\bar{d}_M$.
The potential function is also expressed by the polynomials as in \eqref{VXi}:
\begin{equation}
  V^{\text{KA}}_{\bar{\mathcal{D}}}(x)
  =\kappa^{N+1-M}V(x;\bm{\lambda}-M\bm{\delta})
  \frac{\check{\bar{\Xi}}_{\bar{\mathcal{D}'}}
  (x-i\gamma;\bar{\bm{\lambda}})}
  {\check{\bar{\Xi}}_{\bar{\mathcal{D}'}}(x;\bar{\bm{\lambda}})}
  \frac{\check{\bar{\Xi}}_{\bar{\mathcal{D}}}
  (x+i\frac{\gamma}{2};\bar{\bm{\lambda}})}
  {\check{\bar{\Xi}}_{\bar{\mathcal{D}}}
  (x-i\frac{\gamma}{2};\bar{\bm{\lambda}})}.
  \label{VbarXi}
\end{equation}

The duality between the eigenstates adding and deleting transformations is
stated as the following:
\begin{prop}\label{dual}
For proper parameter ranges in which both Hamiltonians are non-singular
and self-adjoint, the two systems with $\mathcal{H}_{{\mathcal{D}}}$ and
$\mathcal{H}^{\text{\rm KA}}_{\bar{\mathcal{D}}}$ are equivalent.
To be more specific, the equality of the Hamiltonians and the eigenfunctions
read\/{\rm{:}}
\begin{align}
  \mathcal{H}_{{\mathcal{D}}}-\mathcal{E}_{-N-1}(\bm{\lambda})
  &=\kappa^{-N-1}\mathcal{H}^{\text{\rm KA}}_{\bar{\mathcal{D}}},
  \label{Hamdual}\\
  \Phi_{{\mathcal{D}}\,n}(x)
  &\propto\Phi^{\text{\rm KA}}_{\bar{\mathcal{D}}\,n}(x)
  \quad(n=0,1,\ldots),
  \label{eigeq1}\\
  \breve{\Phi}_{{\mathcal{D}};d_j}(x)
  &\propto\breve{\Phi}^{\text{\rm KA}}_{\bar{\mathcal{D}};d_j}(x)
  \quad(j=1,2,\ldots,M).
  \label{eigeq2}
\end{align}
The singularity free conditions of the potential are {\rm \cite{adler,gos}}
\begin{equation}
  \prod_{j=1}^{N+1-M}(n-e_j)\ge0
  \quad(\,\forall n\in\mathbb{Z}_{\geq 0}).
  \label{non-sing}
\end{equation}
\end{prop}
The parameters of the shifted Hamiltonian
$\mathcal{H}^{\text{KA}}_{\bar{\mathcal{D}}}$ are constrained by the
self-adjointness. For the Wilson and Askey-Wilson cases, they are
\begin{equation}
  \text{W}:\ \text{Re}\,a_j>\tfrac12(N+1),\quad
  \text{AW}:\ |a_j|<q^{\frac12(N+1)}\quad(j=1,\ldots,4),
\end{equation}
generalising \eqref{onenarr}.
The two relations \eqref{eigeq1} and \eqref{eigeq2} imply the relationships
among polynomials:
\begin{align}
  P_{\mathcal{D},n}(\eta;\bm{\lambda})&\propto
  \bar{\Xi}_{\bar{\mathcal{D}}\,N+1+n}(\eta;\bar{\bm{\lambda}})
  \quad(n=0,1,\ldots),
  \label{poldual1}\\  
  \Xi_{d_1\ldots\breve{d}_j\ldots d_M}(\eta;\bm{\lambda})&\propto
  \bar{\Xi}_{\bar{\mathcal{D}}\,\bar{d}_j}(\eta;\bar{\bm{\lambda}})
  \quad(j=1,2,\ldots,M).
  \label{poldual2}
\end{align}

%%%%%%%%%%%%%%%%%%%%%%%%%%%%%%%%%%%%%%%%%%%%%%%
%                                             %
% 3.4 Derivation of the Casoratian identities %
%                                             %
%%%%%%%%%%%%%%%%%%%%%%%%%%%%%%%%%%%%%%%%%%%%%%%
\subsection{Derivation of the Casoratian identities}
\label{sec:deriden}

The above duality, {\em i.e.} Proposition \ref{dual}, is the simple
consequence of the following
\begin{prop}\label{polywron}
 The Casoratian Identities read
\begin{equation}
  \Xi_{\mathcal{D}}(\eta;\bm{\lambda})\propto
  \bar{\Xi}_{\bar{\mathcal{D}}}(\eta;\bar{\bm{\lambda}}),
  \label{detiden}
\end{equation}
namely,
\begin{align}
  &\quad\varphi_M(x)^{-1}\,
  \text{\rm W}_{\gamma}[\check{\xi}_{d_1},\check{\xi}_{d_2},\ldots,
  \check{\xi}_{d_M}](x;\bm{\lambda})\n
  &\propto
  \varphi_{N+1-M}(x)^{-1}\,
  \text{\rm W}_{\gamma}[\check{P}_0,\check{P}_1,\ldots,
  \breve{\check{P}}_{\bar{d}_1},\ldots,\breve{\check{P}}_{\bar{d}_M},\ldots,
  \check{P}_N](x;\bar{\bm{\lambda}}).
  \label{genwronide}
\end{align}
\end{prop}
Recall that $\check{\xi}_{\text{v}}(x;\bm{\lambda})=\check{P}_{\text{v}}
\bigl(x;\mathfrak{t}(\bm{\lambda})\bigr)$.
This proposition shows the relation between Casoratians of polynomials
of twisted and shifted parameters.
It is straightforward to show the equality of the Hamiltonians \eqref{Hamdual}
based on the expressions of the potential functions \eqref{VXi}, \eqref{VbarXi}
and $\Xi_{\mathcal{D}}\propto\bar{\Xi}_{\bar{\mathcal{D}}}$ \eqref{detiden}.
The proportionalities of the eigenfunctions \eqref{eigeq1} and \eqref{eigeq2}
follow from the equality of the Hamiltonians, so long as the Hamiltonians
are non-singular and self-adjoint. The inductive proof of
{\bf Proposition \ref{polywron}} in $M$ consists of two steps, as is the
case for the proof of the Wronskian identities in \cite{os29}.

\bigskip
\noindent
\underline{first step} :
As a first step we prove \eqref{detiden} for $M=1$,
$N\ge d_1\equiv \text{v}$, that is $\mathcal{D}=\{\text{v}\}$,
$\bar{\mathcal{D}}=\{0,1,\ldots,\breve{\bar{\text{v}}},\ldots,N\}$
(see Fig.\,2):
\begin{equation}
  \xi_{\text{v}}(\eta;\bm{\lambda})\propto
  \bar{\Xi}_{\bar{\mathcal{D}}}
  (\eta;\bar{\bm{\lambda}}).
  \label{1strel}
\end{equation}

\begin{figure}[htbp]
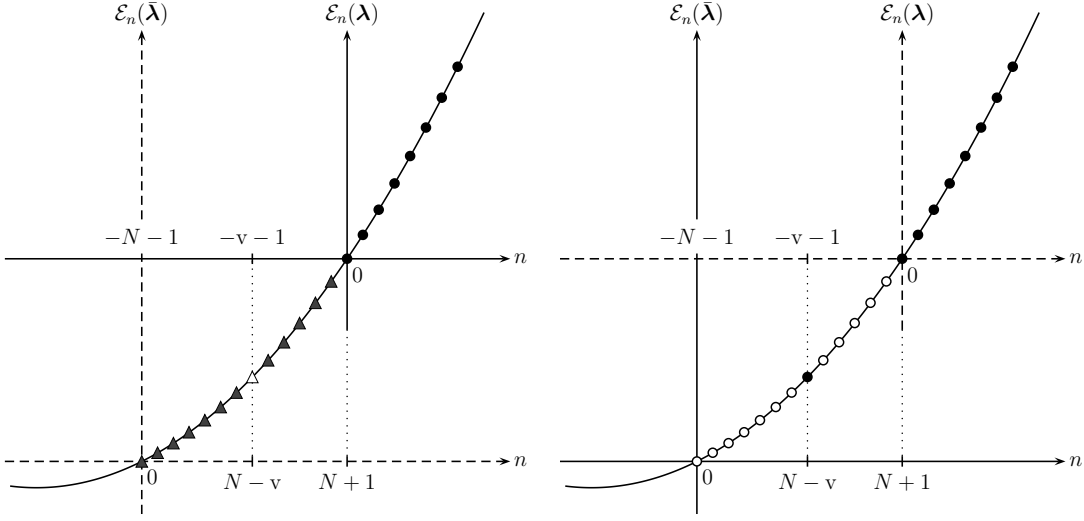

\begin{center}
  \scalebox{0.7}{\includegraphics{enfigIII_pvs2.epsi}}\quad
  \scalebox{0.7}{\includegraphics{enfigIII_es2.epsi}}
  \caption{The symbols are the same as those in Fig.\,1. By shape invariance,
deleting the ground state $\bar{\text v}=N-\text{v}$ from
$\mathcal{H}^{\text{KA}}_{\bar{\mathcal{D}}}$ leads to the undeformed system
$\mathcal{H}(\bm{\lambda})$.
}
\end{center}
\end{figure}

Let us consider a Hamiltonian  system $\bar{\mathcal{H}}$ obtained from
$\mathcal{H}^{\text{KA}}_{\bar{\mathcal{D}}}$ by deleting its ground state
$\bar{\text v}=N-\text{v}$:
\begin{align*}
  &\bar{\mathcal{H}}=\mathcal{A}^{\text{KA}}_{\bar{\mathcal{D}}}
  \mathcal{A}_{\bar{\mathcal{D}}}^{\text{KA}\,\dagger}
  +\mathcal{E}_{\bar{\text{v}}}(\bar{\bm{\lambda}}),\\
  &\bar{\mathcal{H}}\Phi'_n(x)=\mathcal{E}_n(\bar{\bm{\lambda}})\Phi'_n(x)
  \quad(n\geq N+1).
\end{align*}
By shape invariance, the ground state ($n=N+1$) of $\bar{\mathcal{H}}$ coincides
with that of the undeformed system $\mathcal{H}(\bm{\lambda})$, {\em i.e.}
$\phi_0(x;\bm{\lambda})$:
\begin{equation*}
  \bar{\mathcal{H}}\phi_0(x;\bm{\lambda})
  =\mathcal{E}_{N+1}(\bar{\bm{\lambda}})\phi_0(x;\bm{\lambda}).
\end{equation*}
In this case $\bar{\mathcal D}'=\{0,1,\ldots,N\}$ and
$\check{\bar{\Xi}}_{\{0,1,\ldots,N\}}(x;\bar{\bm{\lambda}})=\text{constant}$
\eqref{Xi01..n},
we obtain from \eqref{VbarXi}
\begin{equation*}
  V^{\text{KA}}_{\bar{\mathcal{D}}}(x)=\kappa^NV(x;\bm{\lambda}-\bm{\delta})
  \frac{\check{\bar{\Xi}}_{\bar{\mathcal{D}}}
  (x+i\frac{\gamma}{2};\bar{\bm{\lambda}})}
  {\check{\bar{\Xi}}_{\bar{\mathcal{D}}}
  (x-i\frac{\gamma}{2};\bar{\bm{\lambda}})},
\end{equation*}
and
\begin{align*}
  \bar{\mathcal{H}}&=\kappa^N\Bigl(
  \sqrt{V(x-i\tfrac{\gamma}{2};\bm{\lambda}-\bm{\delta})
  V^*(x-i\tfrac{\gamma}{2};\bm{\lambda}-\bm{\delta})}\,
  e^{\gamma p}\\
  &\phantom{=\kappa^N\Bigl(}
  +\sqrt{V(x+i\tfrac{\gamma}{2};\bm{\lambda}-\bm{\delta})
  V^*(x+i\tfrac{\gamma}{2};\bm{\lambda}-\bm{\delta})}\,
  e^{-\gamma p}\\
  &\phantom{=\kappa^N\Bigl(}
  -V(x+i\tfrac{\gamma}{2};\bm{\lambda}-\bm{\delta})
  \frac{\check{\bar{\Xi}}_{\bar{\mathcal{D}}}
  (x+i\gamma;\bar{\bm{\lambda}})}
  {\check{\bar{\Xi}}_{\bar{\mathcal{D}}}(x;\bar{\bm{\lambda}})}
  -V^*(x-i\tfrac{\gamma}{2};\bm{\lambda}-\bm{\delta})
  \frac{\check{\bar{\Xi}}_{\bar{\mathcal{D}}}
  (x-i\gamma;\bar{\bm{\lambda}})}
  {\check{\bar{\Xi}}_{\bar{\mathcal{D}}}(x;\bar{\bm{\lambda}})}\Bigr)
  +\mathcal{E}_{\bar{\text{v}}}(\bar{\bm{\lambda}}).
\end{align*}
By using the zero mode equation \eqref{zero}, the shape covariance relations
of $\phi_0$ \eqref{phi0shape2}--\eqref{phi0shape3} and of $V$ \eqref{Vshape}
and the general twisting relation \eqref{V'Vs}, we obtain
\begin{align}
  0&=\bigl(\bar{\mathcal{H}}-\mathcal{E}_{N+1}(\bar{\bm{\lambda}})\bigr)
  \phi_0(x;\bm{\lambda})\n
  &=\kappa^{N+1}\Bigl(V(x;\bm{\lambda})+V^*(x;\bm{\lambda})\n
  &\phantom{=\kappa^{N+1}\Bigl(}
  -\alpha(\bm{\lambda})V^*\bigl(x;\mathfrak{t}(\bm{\lambda})\bigr)
  \frac{\check{\bar{\Xi}}_{\bar{\mathcal{D}}}
  (x+i\gamma;\bar{\bm{\lambda}})}
  {\check{\bar{\Xi}}_{\bar{\mathcal{D}}}(x;\bar{\bm{\lambda}})}
  -\alpha(\bm{\lambda})V\bigl(x;\mathfrak{t}(\bm{\lambda})\bigr)
  \frac{\check{\bar{\Xi}}_{\bar{\mathcal{D}}}
  (x-i\gamma;\bar{\bm{\lambda}})}
  {\check{\bar{\Xi}}_{\bar{\mathcal{D}}}(x;\bar{\bm{\lambda}})}
  \Bigr)\phi_0(x;\bm{\lambda})\n
  &\quad+\bigl(\mathcal{E}_{\bar{\text{v}}}(\bar{\bm{\lambda}})
  -\mathcal{E}_{N+1}(\bar{\bm{\lambda}})\bigr)
  \phi_0(x;\bm{\lambda}).
  \label{0=(Hb-}
 \end{align}
With the second basic twist relation \eqref{propV'}, the properties
of $\mathcal{E}_n$ \eqref{enformula1}--\eqref{enformula2} and
$\alpha'(\bm{\lambda})=\mathcal{E}_{-1}(\bm{\lambda})$,
we obtain a difference equation for
$\check{\bar{\Xi}}_{\bar{\mathcal{D}}}(x;\bar{\bm{\lambda}})$:
\begin{align}
  V\bigl(x;\mathfrak{t}(\bm{\lambda})\bigr)
  \bigl(\check{\bar{\Xi}}_{\bar{\mathcal{D}}}(x-i\gamma;\bar{\bm{\lambda}})
  -\check{\bar{\Xi}}_{\bar{\mathcal{D}}}(x;\bar{\bm{\lambda}})\bigr)
  +V^*\bigl(x;\mathfrak{t}(\bm{\lambda})\bigr)
  \bigl(\check{\bar{\Xi}}_{\bar{\mathcal{D}}}(x+i\gamma;\bar{\bm{\lambda}})
  -\check{\bar{\Xi}}_{\bar{\mathcal{D}}}(x;\bar{\bm{\lambda}})\bigr)\n
  =\mathcal{E}_{\text{v}}\bigl(\mathfrak{t}(\bm{\lambda})\bigr)
  \check{\bar{\Xi}}_{\bar{\mathcal{D}}}(x;\bar{\bm{\lambda}}).
  \label{XiDbeq}
\end{align}
This is indeed the difference equation for
$P_{\text{v}}\bigl(\eta(x);\mathfrak{t}(\bm{\lambda})\bigr)$ and we arrive
at the relation
$P_{\text{v}}\bigl(\eta;\mathfrak{t}(\bm{\lambda})\bigr)\propto
\bar{\Xi}_{\bar{\mathcal{D}}}(\eta;\bar{\bm{\lambda}})$ \eqref{1strel}.

\bigskip

\noindent
\underline{second step} :
Assume that \eqref{genwronide} holds till $M$ ($M\ge1$), we will show that
it also holds for $M+1$.

By using the Casoratian identity \eqref{dWformula2}, we obtain
\begin{align*}
  &\quad\text{W}_{\gamma}[\check{\xi}_{d_1},\ldots,\check{\xi}_{d_{M-1}}]
  (x;\bm{\lambda})
  \cdot\text{W}_{\gamma}[\check{\xi}_{d_1},\ldots,\check{\xi}_{d_{M-1}},
  \check{\xi}_{d_M},\check{\xi}_{d_{M+1}}](x;\bm{\lambda})\\
  &=\text{W}_{\gamma}\bigl[
  \text{W}_{\gamma}[\check{\xi}_{d_1},\ldots,\check{\xi}_{d_{M-1}},
  \check{\xi}_{d_M}],
  \text{W}_{\gamma}[\check{\xi}_{d_1},\ldots,\check{\xi}_{d_{M-1}},
  \check{\xi}_{d_{M+1}}]\bigr](x;\bm{\lambda})\\
  &\propto\text{W}_{\gamma}\Bigl[\frac{\varphi_M(x)}{\varphi_{N+1-M}(x)}
  \text{W}_{\gamma}[\check{P}_0,\ldots,\breve{\check{P}}_{\bar{d}_1},\ldots,
  \breve{\check{P}}_{\bar{d}_{M-1}},\ldots,\breve{\check{P}}_{\bar{d}_M},
  \ldots,\check{P}_N],\\
  &\phantom{\propto\text{W}_{\gamma}\Bigl[}
  \ \frac{\varphi_M(x)}{\varphi_{N+1-M}(x)}
  \text{W}_{\gamma}[\check{P}_0,\ldots,\breve{\check{P}}_{\bar{d}_1},\ldots,
  \breve{\check{P}}_{\bar{d}_{M-1}},\ldots,\breve{\check{P}}_{\bar{d}_{M+1}},
  \ldots,\check{P}_N]\Bigr](x;\bar{\bm{\lambda}})\\
  &=\frac{\varphi_M(x+i\frac{\gamma}{2})}{\varphi_{N+1-M}(x+i\frac{\gamma}{2})}
  \frac{\varphi_M(x-i\frac{\gamma}{2})}{\varphi_{N+1-M}(x-i\frac{\gamma}{2})}\n
  &\qquad\times
  \text{W}_{\gamma}\bigl[
  \text{W}_{\gamma}[\check{P}_0,\ldots,\breve{\check{P}}_{\bar{d}_1},\ldots,
  \breve{\check{P}}_{\bar{d}_{M-1}},\ldots,\breve{\check{P}}_{\bar{d}_M},
  \ldots,\check{P}_N],\\
  &\phantom{\qquad\times\text{W}_{\gamma}\bigl[}
  \ \ \text{W}_{\gamma}[\check{P}_0,\ldots,\breve{\check{P}}_{\bar{d}_1},\ldots,
  \breve{\check{P}}_{\bar{d}_{M-1}},\ldots,\breve{\check{P}}_{\bar{d}_{M+1}},
  \ldots,\check{P}_N]\bigr](x;\bar{\bm{\lambda}})\\
  &=\pm
  \frac{\varphi_M(x+i\frac{\gamma}{2})}{\varphi_{N+1-M}(x+i\frac{\gamma}{2})}
  \frac{\varphi_M(x-i\frac{\gamma}{2})}{\varphi_{N+1-M}(x-i\frac{\gamma}{2})}\n
  &\qquad\times
  \text{W}_{\gamma}[\check{P}_0,\ldots,\breve{\check{P}}_{\bar{d}_1},\ldots,
  \breve{\check{P}}_{\bar{d}_{M-1}},\ldots,
  \breve{\check{P}}_{\bar{d}_M},\ldots,
  \breve{\check{P}}_{\bar{d}_{M+1}},\ldots,\check{P}_N](x;\bar{\bm{\lambda}})\\
  &\qquad\times
  \text{W}_{\gamma}[\check{P}_0,\ldots,\breve{\check{P}}_{\bar{d}_1},\ldots,
  \breve{\check{P}}_{\bar{d}_{M-1}},\ldots,\check{P}_N](x;\bar{\bm{\lambda}})\\
  &\propto
  \frac{\varphi_M(x+i\frac{\gamma}{2})}{\varphi_{N+1-M}(x+i\frac{\gamma}{2})}
  \frac{\varphi_M(x-i\frac{\gamma}{2})}{\varphi_{N+1-M}(x-i\frac{\gamma}{2})}
  \frac{\varphi_{N+2-M}(x)}{\varphi_{M-1}(x)}\n
  &\qquad\times
  \text{W}_{\gamma}[\check{P}_0,\ldots,\breve{\check{P}}_{\bar{d}_1},\ldots,
  \breve{\check{P}}_{\bar{d}_{M+1}},\ldots,\check{P}_N](x;\bar{\bm{\lambda}})
  \cdot\text{W}_{\gamma}[\check{\xi}_{d_1},\ldots,\check{\xi}_{d_{M-1}}]
  (x;\bm{\lambda}).
\end{align*}
This leads to
\begin{align*}
  &\quad\varphi_{M+1}(x)^{-1}\,
  \text{W}_{\gamma}[\check{\xi}_{d_1},\ldots,\check{\xi}_{d_{M+1}}]
  (x;\bm{\lambda})\n
  &\propto
  \varphi_{N-M}(x)^{-1}\,
  \text{W}_{\gamma}[\check{P}_0,\ldots,\breve{\check{P}}_{\bar{d}_1},\ldots,
  \breve{\check{P}}_{\bar{d}_{M+1}},\ldots,\check{P}_N](x;\bar{\bm{\lambda}})\n
  &\qquad\times
  \frac{\varphi_M(x+i\frac{\gamma}{2})}{\varphi_{N+1-M}(x+i\frac{\gamma}{2})}
  \frac{\varphi_M(x-i\frac{\gamma}{2})}{\varphi_{N+1-M}(x-i\frac{\gamma}{2})}
  \frac{\varphi_{N+2-M}(x)}{\varphi_{M-1}(x)}
  \frac{\varphi_{N-M}(x)}{\varphi_{M+1}(x)}\n
  &\propto
  \varphi_{N-M}(x)^{-1}\,
  \text{W}_{\gamma}[\check{P}_0,\ldots,\breve{\check{P}}_{\bar{d}_1},\ldots,
  \breve{\check{P}}_{\bar{d}_{M+1}},\ldots,\check{P}_N](x;\bar{\bm{\lambda}}),
\end{align*}
namely $M+1$ case is shown.
Here we have used
\begin{equation}
  \varphi_M(x+i\tfrac{\gamma}{2})\varphi_M(x-i\tfrac{\gamma}{2})
  =\frac{\varphi_{M-1}(x)\varphi_{M+1}(x)}{\varphi(x)}
  \quad(M\geq 1),
\end{equation}
which is easily verified.
This concludes the induction proof of \eqref{genwronide}.

%%%%%%%%%%%%%%%%%%%%%%%%%%%%%%%%%%%%%%%%%%%%%%%%%%%%%%%%%%%%%%%
%                                                             %
%  4. Reduced Case Polynomials                                %
%                                                             %
%%%%%%%%%%%%%%%%%%%%%%%%%%%%%%%%%%%%%%%%%%%%%%%%%%%%%%%%%%%%%%%
\section{Reduced Case Polynomials}
\label{sec:red}

It is well known that the other members of the Askey scheme polynomials can
be obtained by reductions from the Wilson and the Askey-Wilson polynomials.
Here we list the discrete symmetry transformations and the pseudo virtual
state wave functions for all the reduced case polynomials.
In contrast to the virtual state wave functions, the pseudo virtual state
wave functions are universal and they exist for all the solvable potentials
with shape invariance.
For example, for the systems of the harmonic oscillator and the $q$-harmonic
oscillator with the ($q$-)Hermite polynomials as the main part of the
eigenfunctions \cite{os11}, virtual state wave functions do not exist.
However, the pseudo virtual state wave functions for the harmonic oscillator
was reported in \cite{os29} and those for the $q$-harmonic oscillator will
be introduced in \S\,\ref{sec:redAW}.
The Casoratian identities hold for these reduced case polynomials, too.

%%%%%%%%%%%%%%%%%%%%%%%%%%%%%%%%%%%%%%%%%%%%%
%                                           %
% 4.1 Reductions from the Wilson polynomial %
%                                           %
%%%%%%%%%%%%%%%%%%%%%%%%%%%%%%%%%%%%%%%%%%%%%
\subsection{Reductions from the Wilson polynomial}
\label{sec:redW}

Three polynomials belong to this group; the continuous dual Hahn (cdH),
the continuous Hahn (cH) and the Meixner-Pollaczek (MP) polynomials.
They are obtained from the Wilson polynomial by some limiting procedures
\cite{koeswart}. The discrete symmetries are also obtained by the same
limiting procedures from that of the Wilson polynomial.
It should be noted that the Wilson polynomial is also obtained from the
Askey-Wilson polynomial by a certain limiting procedure \cite{koeswart}.

The defining domain and the parameters of these reduced case polynomials are:
\begin{alignat}{7}
  \text{cdH}:\ \ &x_1=0,\ \ &x_2=\infty,\quad&\gamma=1,
  &\quad&\bm{\lambda}=(a_1,a_2,a_3),
  &\quad&\bm{\delta}=(\tfrac12,\tfrac12,\tfrac12),
  \quad&\kappa=1,\n
  \text{cH}:\ \ &x_1=-\infty,\ \ &x_2=\infty,\quad&\gamma=1,
  &\quad&\bm{\lambda}=(a_1,a_2),
  &\quad&\bm{\delta}=(\tfrac12,\tfrac12),
  \quad&\kappa=1,\n
  \text{MP}:\ \ &x_1=-\infty,\ \ &x_2=\infty,\quad&\gamma=1,
  &\quad&\bm{\lambda}=(a,\phi),
  &\quad&\bm{\delta}=(\tfrac12,0),
  \quad&\kappa=1,
\end{alignat}
in which the parameters are restricted by
\begin{align}
  \text{cdH}:&\ \ \{a_1^*,a_2^*,a_3^*\}=\{a_1,a_2,a_3\}
  \ \ (\text{as a set});\quad\text{Re}\,a_i>0,\n
  \text{cH}:&\ \ \text{Re}\,a_i>0;\quad(a_3,a_4)\eqdef(a_1^*,a_2^*),\n
  \text{MP}:&\ \ a>0,\quad 0<\phi<\pi.
  \end{align}
Here are the fundamental data:
\begin{align}
  V(x;\bm{\lambda})&=\left\{
  \begin{array}{ll}
  \bigl(2ix(2ix+1)\bigr)^{-1}\prod_{j=1}^3(a_j+ix)&:\text{cdH}\\[4pt]
  \prod_{j=1}^2(a_j+ix)&:\text{cH}\\[4pt]
  e^{i(\frac{\pi}{2}-\phi)}(a+ix)&:\text{MP}
  \end{array}\right.,\\
  \eta(x)&=\left\{
  \begin{array}{ll}
  x^2&:\text{cdH}\\
  x&:\text{cH,\,MP}
  \end{array}\right.,\quad
  \varphi(x)=\left\{
  \begin{array}{ll}
  2x&:\text{cdH}\\
  1&:\text{cH,\,MP}
  \end{array}\right.,
  \label{etadef2}\\
  \mathcal{E}_n(\bm{\lambda})&=\left\{
  \begin{array}{ll}
  n&:\text{cdH}\\
  n(n+b_1-1),\quad
  b_1\eqdef a_1+a_2+a_3+a_4,&:\text{cH}\\[2pt]
  2n\sin\phi&:\text{MP}
  \end{array}\right.,\\
  \phi_n(x;\bm{\lambda})
  &=\phi_0(x;\bm{\lambda})\check{P}_n(x;\bm{\lambda}),\\
  \check{P}_n(x;\bm{\lambda})&=P_n\bigl(\eta(x);\bm{\lambda}\bigr)
  =\left\{
  \begin{array}{ll}
  S_n\bigl(\eta(x);a_1,a_2,a_3\bigr)&:\text{cdH}\\[1pt]
  p_n\bigl(\eta(x);a_1,a_2,a_3,a_4\bigr)&:\text{cH}\\[2pt]
  P_n^{(a)}\bigl(\eta(x);\phi\bigr)&:\text{MP}
  \end{array}\right.\n
  &=\left\{
  \begin{array}{ll}
  {\displaystyle
  (a_1+a_2,a_1+a_3)_n
  \ {}_3F_2\Bigl(\genfrac{}{}{0pt}{}{-n,\,a_1+ix,\,a_1-ix}
  {a_1+a_2,\,a_1+a_3}\Bigm|1\Bigr)}
  &:\text{cdH}\\[8pt]
  {\displaystyle
  i^n\frac{(a_1+a_3,a_1+a_4)_n}{n!}
  \ {}_3F_2\Bigl(\genfrac{}{}{0pt}{}{-n,\,n+b_1-1,\,a_1+ix}
  {a_1+a_3,\,a_1+a_4}\!\Bigm|\!1\Bigr)}
  &:\text{cH}\\[8pt]
  {\displaystyle 
  \frac{(2a)_n}{n!}\,e^{in\phi}
  {}_2F_1\Bigl(\genfrac{}{}{0pt}{}{-n,\,a+ix}{2a}\Bigm|
  1-e^{-2i\phi}\Bigr)}
  &:\text{MP}
  \end{array}\right.,
  \label{Pn=cdH}\\[4pt]
  \phi_0(x;\bm{\lambda})&=\left\{
  \begin{array}{ll}
  \sqrt{(\Gamma(2ix)\Gamma(-2ix))^{-1}\prod_{j=1}^3
  \Gamma(a_j+ix)\Gamma(a_j-ix)}
  &:\text{cdH}\\[6pt]
  \sqrt{\Gamma(a_1+ix)\Gamma(a_2+ix)\Gamma(a_3-ix)\Gamma(a_4-ix)}
  &:\text{cH}\\[6pt]
  e^{(\phi-\frac{\pi}{2})x}\sqrt{\Gamma(a+ix)\Gamma(a-ix)}
  &:\text{MP}
  \end{array}\right.,\\
  &f_n(\bm{\lambda})=\left\{
  \begin{array}{ll}
  -n&:\text{cdH}\\
  n+b_1-1&:\text{cH}\\
  2\sin\phi&:\text{MP}
  \end{array}\right.,\quad
  b_{n-1}(\bm{\lambda})=\left\{
  \begin{array}{ll}
  -1&:\text{cdH}\\
  n&:\text{cH}\\
  n&:\text{MP}
  \end{array}\right..
\end{align}
The relations \eqref{Vshape}--\eqref{phi0shape} are satisfied.

%%%%%%%%%%%%%%%%%%%%%%%%%%%%%%%%%%%%%%%%%%%%%
%                                           %
% 4.1.1 pseudo virtual state wave functions %
%                                           % 
%%%%%%%%%%%%%%%%%%%%%%%%%%%%%%%%%%%%%%%%%%%%%
\subsubsection{pseudo virtual state wave functions}
\label{sec:pseudoW}

The twisting of the polynomials in this group is straightforward.
We define the twisted potential $V'(x;\bm{\lambda})$ \eqref{V'def1} by
\begin{align}
  \mathfrak{t}(\bm{\lambda})&=\left\{
  \begin{array}{ll}
  (1-a_1,1-a_2,1-a_3)&:\text{cdH}\\
  (1-a_1^*,1-a_2^*)&:\text{cH}\\
  (1-a,\pi-\phi)&:\text{MP}\\
  \end{array}\right.,\quad
  \mathfrak{t}^2=\text{Id}.
\end{align}
The relations \eqref{VV'rel}--\eqref{propV'}, \eqref{vminv} and \eqref{V'Vs}
are satisfied with
\begin{align}
\alpha(\bm{\lambda})=\left\{
  \begin{array}{ll}
  -1&:\text{cdH,\,MP}\\
  1&:\text{cH}
  \end{array}\right.,\qquad
  \alpha'(\bm{\lambda})=\mathcal{E}_{-1}(\bm{\lambda})=\left\{
  \begin{array}{ll}
  -1&:\text{cdH}\\
  2-b_1&:\text{cH}\\
  -2\sin\phi &:\text{MP}
  \end{array}\right., 
\end{align}
and the pseudo virtual state wave function is obtained by simple twisting of
the parameters
$\tilde{\phi}_{\text{v}}(x;\bm{\lambda})
=\tilde{\phi}_0(x;\bm{\lambda})\check{\xi}_ {\text{v}}(x;\bm{\lambda})$
as in \eqref{psvfac}--\eqref{psvfac2}.

%%%%%%%%%%%%%%%%%%%%%%%%%%%%%%%%%%%%%%%%%%%%%%%%%%%
%                                                 % 
% 4.2 Reductions from the Askey-Wilson polynomial %
%                                                 %
%%%%%%%%%%%%%%%%%%%%%%%%%%%%%%%%%%%%%%%%%%%%%%%%%%% 
\subsection{Reductions from the Askey-Wilson polynomial}
\label{sec:redAW}

There are two groups, to be called (A) and (B), of polynomials obtained
by two different types of reductions from the Askey-Wilson polynomial.
Group (A), consisting of one polynomial, is obtained by specifying the
four parameters $(a_1,a_2,a_3,a_4)$ of the Askey-Wilson polynomial,
as simple functions of two ($\alpha$ and $\beta$) parameters.
Group (B), consisting of five polynomials, is obtained by setting some of
the parameters $\{a_j\}$ to zero.
For all member polynomials in this subsection, we have
\begin{equation*}
  x_1=0,\quad x_2=\pi,\quad\gamma=\log q,\quad\kappa=q^{-1},\quad
  \eta(x)=\cos x,\quad\varphi(x)=2\sin x.
\end{equation*}

%%%%%%%%%%%%%%%%%%%%%%%%%%%%%%%%%%%%%%%%%%%%%%%%%%%%%%%%%%%%%%%
%                                                             % 
% 4.2.1 Group (A) reductions from the Askey-Wilson polynomial %
%                                                             %
%%%%%%%%%%%%%%%%%%%%%%%%%%%%%%%%%%%%%%%%%%%%%%%%%%%%%%%%%%%%%%%
\subsubsection{Group (A) reductions from the Askey-Wilson polynomial}
\label{sec:redA}

The continuous $q$-Jacobi (c$q$J) polynomial belongs to this group.
It is obtained by restricting the four parameters $(a_1,a_2,a_3,a_4)$ of
the Askey-Wilson polynomial as
\begin{align}
  &(a_1, a_2, a_3, a_4)
  =\bigl(q^{\frac12(\alpha+\frac12)},q^{\frac12(\alpha+\frac32)},
  -q^{\frac12(\beta+\frac12)},-q^{\frac12(\beta+\frac32)}\bigr),
  \label{cqJdef}\\
  &\bm{\lambda}=(\alpha,\beta),\quad \bm{\delta}=(1,1),\quad
  \alpha,\beta\ge-\tfrac12,\\
  &V(x\,;\bm{\lambda})=
  \frac{(1-q^{\frac12(\alpha+\frac12)}e^{ix})(1-q^{\frac12(\alpha+\frac32)}
  e^{ix})
  (1+q^{\frac12(\beta+\frac12)}e^{ix})(1+q^{\frac12(\beta+\frac32)}e^{ix})}
  {(1-e^{2ix})(1-qe^{2ix})}.
\end{align}
The eigenvalues and the corresponding eigenfunctions are:
\begin{align}
  \mathcal{E}_n(\bm{\lambda})&=(q^{-n}-1)(1-q^{n+\alpha+\beta+1}),\\
  \check{P}_n(x;\bm{\lambda})&=P_n\bigl(\eta(x);\bm{\lambda}\bigr)
  =P_n^{(\alpha,\beta)}\bigl(\eta(x)|q\bigr)\n
  &=\frac{(q^{\alpha+1}\,;q)_n}{(q\,;q)_n}\,
  {}_4\phi_3\Bigl(\genfrac{}{}{0pt}{}{q^{-n},\,q^{n+\alpha+\beta+1},\,
  q^{\frac12(\alpha+\frac12)}e^{ix},\,q^{\frac12(\alpha+\frac12)}e^{-ix}}
  {q^{\alpha+1},\,-q^{\frac12(\alpha+\beta+1)},\,
  -q^{\frac12(\alpha+\beta+2)}}\Bigm|q\,;q\Bigr),
  \label{defcqJ}\\[4pt]
  \phi_0(x\,;\bm{\lambda})&=
  \sqrt{\frac{(e^{2ix},e^{-2ix}\,;q)_{\infty}}
  {(q^{\frac12(\alpha+\frac12)}e^{ix},
  -q^{\frac12(\beta+\frac12)}e^{ix},
  q^{\frac12(\alpha+\frac12)}e^{-ix},
  -q^{\frac12(\beta+\frac12)}e^{-ix}\,;q^{\frac12})_{\infty}}},
  \label{qJphi0}\\
  f_n(\bm{\lambda})&=\frac{q^{\frac12(\alpha+\frac32)}q^{-n}
  (1-q^{n+\alpha+\beta+1})}
  {(1+q^{\frac12(\alpha+\beta+1)})(1+q^{\frac12(\alpha+\beta+2)})},\\
  b_{n-1}(\bm{\lambda})&=q^{-\frac12(\alpha+\frac32)}q^n(q^{-n}-1)
  (1+q^{\frac12(\alpha+\beta+1)})(1+q^{\frac12(\alpha+\beta+2)}).
\end{align}
The relations \eqref{Vshape}--\eqref{phi0shape} are satisfied.

%%%%%%%%%%%%%%%%%%%%%%%%%%%%%%%%%%%%%%%%%%%%%
%                                           % 
% 4.2.2 pseudo virtual states for Group (A) %
%                                           %
%%%%%%%%%%%%%%%%%%%%%%%%%%%%%%%%%%%%%%%%%%%%%
\subsubsection{pseudo virtual states for Group (A)}
\label{sec:pseudoA}

The twisting of the Askey-Wilson case \eqref{WAWtwist} is consistent with
the reduction to Group (A). That is $a_j\to q\,a_j^{-1}$ ($j=1,\ldots,4$)
simply translates to the twisting of the two parameters $\alpha$ and $\beta$:
\begin{align}
  \mathfrak{t}(\alpha,\beta)=(-\alpha,-\beta),\quad\mathfrak{t}^2=\text{Id},
\end{align}
giving the potential $V'(x;\bm{\lambda})$ \eqref{V'def1}.
The relations \eqref{VV'rel}--\eqref{propV'}, \eqref{vminv} and
\eqref{V'Vs} are satisfied  with
\begin{align} 
  \alpha(\bm{\lambda})=q^{\alpha+\beta},\quad
  \alpha'(\bm{\lambda})=\mathcal{E}_{-1}(\bm{\lambda})=
  (q-1)(1-q^{\alpha+\beta}),
\end{align}
and the pseudo virtual state wave function is obtained by simple twisting
of the parameters
$\tilde{\phi}_{\text{v}}(x;\bm{\lambda})=\tilde{\phi}_0(x;\bm{\lambda})
\check{\xi}_{\text{v}}(x;\bm{\lambda})$ as in \eqref{psvfac}--\eqref{psvfac2}.

%%%%%%%%%%%%%%%%%%%%%%%%%%%%%%%%%%%%%%%%%%%%%%%%%%%%%%%%%%%%%%%
%                                                             %
% 4.2.3 Group (B) reductions from the Askey-Wilson polynomial %
%                                                             %
%%%%%%%%%%%%%%%%%%%%%%%%%%%%%%%%%%%%%%%%%%%%%%%%%%%%%%%%%%%%%%%
\subsubsection{Group (B) reductions from the Askey-Wilson polynomial}
\label{sec:redB}

Five polynomials belong to Group (B); the continuous dual $q$-Hahn (cd$q$H),
Al-Salam-Chihara (ASC), continuous big $q$-Hermite (cb$q$H), continuous
$q$-Hermite (c$q$H) and continuous $q$-Laguerre (c$q$L) polynomials.
The first four members are obtained by setting $a_4=0$ for cd$q$H,
$a_4=a_3=0$ for ASC, $a_4=a_3=a_2=0$ for cb$q$H and $a_4=a_3=a_2=a_1=0$
for c$q$H. The c$q$L is obtained by setting $a_4=a_3=0$ of the continuous
$q$-Jacobi case. In other words, the c$q$L is obtained by taking the
limit $\beta\to+\infty$ of the continuous $q$-Jacobi case.
The parameters of Group (B) are
\begin{alignat}{6}
  \text{cd$q$H}:\ \ &\bm{\lambda}=(\lambda_1,\lambda_2,\lambda_3),\quad
  &&q^{\bm{\lambda}}=(a_1,a_2,a_3),\quad
  &&\bm{\delta}=(\tfrac12,\tfrac12,\tfrac12),\quad&&|a_j|<1,\\
  \text{ASC}:\ \ &\bm{\lambda}=(\lambda_1,\lambda_2),\quad
  &&q^{\bm{\lambda}}=(a_1,a_2),\quad
  &&\bm{\delta}=(\tfrac12,\tfrac12),\quad&&|a_j|<1,\\
  \text{cb$q$H}:\ \ &\bm{\lambda}=\lambda_1,\quad
  &&q^{\bm{\lambda}}=a_1=a,\quad
  &&\bm{\delta}=\tfrac12,\quad&&|a|<1,\\
  \text{c$q$H}:\ \ &\bm{\lambda}:\ \text{none},\\
  \text{c$q$L}:\ \ &\bm{\lambda}=\alpha,
  &&&&\bm{\delta}=1,\quad&&\alpha>-\tfrac12.
\end{alignat}
The basic data are obtained from those of the Askey-Wilson and the continuous
$q$-Jacobi polynomials by simply putting the appropriate parameters to zero:
\begin{align}
  &V(x;\bm{\lambda},q)=\frac{1}{(1-e^{2ix})(1-qe^{2ix})}
  \times\left\{
  \begin{array}{l}
  \prod_{j=1}^m(1-a_je^{ix})
  \qquad\qquad:m=3,2,1,0\\[6pt]
  (1-q^{\frac12(\alpha+\frac12)}e^{ix})(1-q^{\frac12(\alpha+\frac32)}e^{ix})
  \ \ :\text{c$q$L}
  \end{array}\right.,
  \label{Vformreds}\\
  &\mathcal{E}_n(\bm{\lambda})=q^{-n}-1,\\
  &\phi_0(x;\bm{\lambda})=
  \sqrt{(e^{2ix},e^{-2ix}\,;q)_{\infty}}
  \times\left\{
  \begin{array}{l}
  \sqrt{\prod_{j=1}^m(a_je^{ix},a_je^{-ix}\,;q)_{\infty}^{-1}}
  \quad\ :m=3,2,1,0\\[6pt]
  \sqrt{(q^{\frac12(\alpha+\frac12)}e^{ix},
  q^{\frac12(\alpha+\frac12)}e^{-ix}\,;q^{\frac12})_{\infty}^{-1}}
  \ \ :\text{c$q$L}
  \end{array}\right.,\\
  &\check{P}_n(x;\bm{\lambda})=P_n\bigl(\eta(x);\bm{\lambda}\bigr)=\left\{
  \begin{array}{ll}
  p_n\bigl(\eta(x);a_1,a_2,a_3|q\bigr)
  &:\text{cd$q$H}\\
  Q_n\bigl(\eta(x);a_1,a_2|q\bigr)
  &:\text{ASC}\\
  H_n\bigl(\eta(x);a|q\bigr)
  &:\text{cb$q$H}\\
  H_n\bigl(\eta(x)|q\bigr)
  &:\text{c$q$H}\\
  P_n^{(\alpha)}\bigl(\eta(x)|q\bigr)
  &:\text{c$q$L}
  \end{array}\right.\n
  &\phantom{\check{P}_n(x;\bm{\lambda})}
  =\left\{
  \begin{array}{ll}
  {\displaystyle
  a_1^{-n}(a_1a_2,a_1a_3\,;q)_n\,
  {}_3\phi_2\Bigl(\genfrac{}{}{0pt}{}{q^{-n},\,
  a_1e^{ix},\,a_1e^{-ix}}{a_1a_2,\,a_1a_3}\Bigm|q\,;q\Bigr)}
  &:\text{cd$q$H}\\[6pt]
  {\displaystyle
  a_1^{-n}(a_1a_2\,;q)_n\,
  {}_3\phi_2\Bigl(\genfrac{}{}{0pt}{}{q^{-n},\,
  a_1e^{ix},\,a_1e^{-ix}}{a_1a_2,\,0}\Bigm|q\,;q\Bigr)}
  &:\text{ASC}\\[6pt]
  {\displaystyle
  a^{-n}\,
  {}_3\phi_2\Bigl(\genfrac{}{}{0pt}{}{q^{-n},\,
  ae^{ix},\,ae^{-ix}}{0,\,0}\Bigm|q\,;q\Bigr)}
  &:\text{cb$q$H}\\[6pt]
  {\displaystyle
  e^{inx}\,
  {}_2\phi_0\Bigl(\genfrac{}{}{0pt}{}{q^{-n},\,0}{-}
  \Bigm|q\,;q^ne^{-2ix}\Bigr)}
  &:\text{c$q$H}\\[6pt]
  {\displaystyle
  \frac{(q^{\alpha+1}\,;q)_n}{(q\,;q)_n}\,
  {}_3\phi_2\Bigl(\genfrac{}{}{0pt}{}{q^{-n},\,
  q^{\frac12(\alpha+\frac12)}e^{ix},\,q^{\frac12(\alpha+\frac12)}e^{-ix}}
  {q^{\alpha+1},\,0}\Bigm|q\,;q\Bigr)}
  &:\text{c$q$L}
  \end{array}\right.,\\
  &f_n(\bm{\lambda})=\left\{
  \begin{array}{ll}
  q^{\frac{n}{2}}(q^{-n}-1)&:\text{cd$q$H,\,ASC,\,cb$q$H,\,c$q$H}\\[2pt]
  q^{\frac12(\alpha+\frac32)}q^{-n}&:\text{c$q$L}
  \end{array}\right.,\\
  &b_{n-1}(\bm{\lambda})=\left\{
  \begin{array}{ll}
  q^{-\frac{n}{2}}&:\text{cd$q$H,\,ASC,\,cb$q$H,\,c$q$H}\\
  q^{-\frac12(\alpha+\frac32)}q^n(q^{-n}-1)&:\text{c$q$L}
  \end{array}\right.,
\end{align}
where $m=3,2,1,0$ correspond to cd$q$H, ASC, cb$q$H, c$q$H, respectively.
The relations \eqref{Vshape}--\eqref{phi0shape} are satisfied.

%%%%%%%%%%%%%%%%%%%%%%%%%%%%%%%%%%%%%%%%%%%%%
%                                           %
% 4.2.4 pseudo virtual states for Group (B) %
%                                           %
%%%%%%%%%%%%%%%%%%%%%%%%%%%%%%%%%%%%%%%%%%%%%
\subsubsection{pseudo virtual states for Group (B)}
\label{sec:pseudoB}

The twisting of the Askey-Wilson case \eqref{WAWtwist} is {\em not\/}
consistent with the reduction to Group (B). As can be seen clearly the
transformation $a_j\to q\,a_j^{-1}$ ($j=1,2,3$) in cd$q$H potential
$V(x;\bm{\lambda})$ simply fails to satisfy the two basic relations
\eqref{VV'rel} and \eqref{propV'}. For the c$q$H, having no parameter
other than $q$, such a transformation using the twisting of $a_j$ is
simply meaningless.

As can be easily guessed, the desired twisting should include the twisting
of the parameter $q$ as its part, if it should cover the c$q$H case.
We write $q$-dependence explicitly, if necessary.
We propose the following twisting:
\begin{align}
  &V'(x;\bm{\lambda})\eqdef V\bigl(-x;\mathfrak{t}(\bm{\lambda}),q^{-1}\bigr)
  =V^*\bigl(x;\mathfrak{t}(\bm{\lambda}),q^{-1}\bigr),\\
  &\mathfrak{t}(\bm{\lambda})=\left\{
  \begin{array}{lll}
  (1-\lambda_1,1-\lambda_2,1-\lambda_3)&:\text{cd$q$H}
  &\bigl(\ \text{or}\ \ a_j\to a_jq^{-1}\ \ (j=1,2,3)\,\bigr)\\
  (1-\lambda_1,1-\lambda_2)&:\text{ASC}
  &\bigl(\ \text{or}\ \ a_j\to a_jq^{-1}\ \ (j=1,2)\,\bigr)\\
  1-\lambda_1&:\text{cb$q$H}
  &\bigl(\ \text{or}\ \ a\ \to a\,q^{-1}\ \bigr)\\
  \text{none}&:\text{c$q$H}\\
  -\alpha&:\text{c$q$L}
  \end{array}\right.,
\end{align}
which satisfies the relations \eqref{VV'rel}--\eqref{propV'},
\eqref{vminv} and \eqref{V'Vs} with
\begin{align}
  &\tilde{\mathcal{E}}_{\text{v}}(\bm{\lambda})
  \eqdef\alpha(\bm{\lambda})
  \mathcal{E}_{\text{v}}\bigl(\mathfrak{t}(\bm{\lambda}),q^{-1}\bigr)
  +\alpha'(\bm{\lambda})
  =\mathcal{E}_{-\text{v}-1}(\bm{\lambda}),\n
  &\alpha(\bm{\lambda})=q,\quad
  \alpha'(\bm{\lambda})=\mathcal{E}_{-1}(\bm{\lambda})=q-1,
\end{align}
for every member polynomial in Group (B).
Note that $V''(x;\bm{\lambda})=V(x;\bm{\lambda})$.

The corresponding pseudo virtual state wave functions are given by
\begin{align}
  \tilde{\phi}_{\text{v}}(x;\bm{\lambda})
  &=\tilde{\phi}_0(x;\bm{\lambda})\check{\xi}_{\text{v}}(x;\bm{\lambda}),\\
  \tilde{\phi}_0(x;\bm{\lambda})
  &\eqdef\frac{\varphi(x)}{\phi_0(x;\bm{\lambda})},\quad
  \check{\xi}_{\text{v}}(x;\bm{\lambda})
  \eqdef\check{P}_{\text{v}}\bigl(x;\mathfrak{t}(\bm{\lambda}),q^{-1}\bigr)
  =P_{\text{v}}\bigl(\eta(x);\mathfrak{t}(\bm{\lambda}),q^{-1}\bigr).
  \label{newzero}
\end{align}
It should be stressed that the above zero mode of
$\mathcal{A}'(\bm{\lambda})$, $\tilde{\phi}_0(x;\bm{\lambda})$, is not
obtained by replacing $q\to q^{-1}$ and
$\bm{\lambda}\to \mathfrak{t}(\bm{\lambda})$ in the original zero mode
$\phi_0(x;\bm{\lambda})$, since infinite products like $(e^{2ix};q)_\infty$
contained in $\phi_0(x;\bm{\lambda})$ do not converge if $q$ is replaced by
$q^{-1}$. The above form \eqref{newzero} of the zero mode is obtained from
the linear relation \eqref{V'Vs} between the twisted potential and the
original potential:
\begin{align*}
  \eqref{V'Vs}\Rightarrow\quad&
  V'(x+i\tfrac{\gamma}{2};\bm{\lambda})
  =\alpha(\bm{\lambda})^{-1}
  \frac{\varphi(x-i\frac{\gamma}{2})}{\varphi(x+i\frac{\gamma}{2})}
  V^*(x-i\tfrac{\gamma}{2};\bm{\lambda}),\\
  &\!\!\!V^{\prime\,*}(x-i\tfrac{\gamma}{2};\bm{\lambda})
  =\alpha(\bm{\lambda})^{-1}
  \frac{\varphi(x+i\frac{\gamma}{2})}{\varphi(x-i\frac{\gamma}{2})}
  V(x+i\tfrac{\gamma}{2};\bm{\lambda}).
\end{align*}
Then the zero mode equation
\begin{equation*}
  \sqrt{V^{\prime\,*}(x-i\tfrac{\gamma}{2};\bm{\lambda})}
  \,\tilde{\phi}_0(x-i\tfrac{\gamma}{2};\bm{\lambda})
  =\sqrt{V'(x+i\tfrac{\gamma}{2};\bm{\lambda})}
  \,\tilde{\phi}_0(x+i\tfrac{\gamma}{2};\bm{\lambda})
\end{equation*}
can be rewritten as
\begin{equation*}
  \sqrt{V^*(x-i\tfrac{\gamma}{2};\bm{\lambda})}\,
  \varphi(x-i\tfrac{\gamma}{2})
  \tilde{\phi}_0(x-i\tfrac{\gamma}{2};\bm{\lambda})^{-1}
  =\sqrt{V\bigl(x+i\tfrac{\gamma}{2};\bm{\lambda})}\,
  \varphi(x+i\tfrac{\gamma}{2})
  \tilde{\phi}_0(x+i\tfrac{\gamma}{2};\bm{\lambda})^{-1},
\end{equation*}
which simply means \eqref{newzero},
$\varphi(x)\tilde{\phi}_0(x;\bm{\lambda})^{-1}={\phi}_0(x;\bm{\lambda})$.

For the Askey-Wilson polynomial $p_n(\eta)$ \eqref{Pn=W,AW}, it is possible
to twist as a member of Group (B).
The two types of twisted polynomials are proportional to each other
\cite{koeswart}
\begin{align}
  &p_n\bigl(\eta;qa_1^{-1},qa_2^{-1},qa_3^{-1},qa_4^{-1}|\,q\bigr)\n
  &\quad=(-1)^n\frac{q^{\frac12n(3n+5)}}{(a_1a_2a_3a_4)^n}
  p_n\bigl(\eta;a_1q^{-1},a_2q^{-1},a_3q^{-1},a_4q^{-1}|\,q^{-1}\bigr),
  \label{AWqi}
\end{align}
and they lead to the same deformation.
Similar relation holds for  the continuous $q$-Jacobi polynomial.

%%%%%%%%%%%%%%%%%%%%%%%%%%%%%%%%%%%%%%%%%%%%%%%%%%%%%%%%%
%                                                       %
% 4.3 Casoratian identities for the reduced polynomials %
%                                                       %
%%%%%%%%%%%%%%%%%%%%%%%%%%%%%%%%%%%%%%%%%%%%%%%%%%%%%%%%% 
\subsection{Casoratian identities for the reduced polynomials}
\label{sec:redCas}

Casoratian identities also hold for the reduced case polynomials.
The derivation in \S\,\ref{sec:deriden} is valid for them
(eqs.\eqref{0=(Hb-}--\eqref{XiDbeq} should be slightly modified).
The necessary formulas are \eqref{Vshape}--\eqref{phi0shape},
\eqref{VV'rel}--\eqref{propV'}, \eqref{V'Vs} and the properties of
$\mathcal{E}_n$ (\eqref{enformula1}--\eqref{enformula2}, \eqref{vminv}
and $\alpha'(\bm{\lambda})=\mathcal{E}_{-1}(\bm{\lambda})$).
Definitions of the various quantities such as $A_{\mathcal{D}}$,
$\Xi_{\mathcal{D}}$, $A_{\mathcal{D},n}$, $P_{\mathcal{D},n}$, $\nu$,
$r_j$ etc. are the same.
The explicit forms of $r_j$ for the reduced polynomials in \S\,\ref{sec:redW}
are
\begin{align}
  &\quad r_j(x^{(M+1)}_j;\bm{\lambda},M+1)\n
  &\quad\propto\left\{
  \begin{array}{ll}
  \prod_{k=1}^3(a_k-\frac{M}{2}+ix)_{j-1}(a_k-\frac{M}{2}-ix)_{M+1-j}
  &:\text{cdH}\\[4pt]
  \prod_{k=1}^2(a_k-\frac{M}{2}+ix)_{j-1}(a_k^*-\frac{M}{2}-ix)_{M+1-j}
  &:\text{cH}\\[4pt]
  e^{2i(\phi-\frac{\pi}{2})(\frac{M}{2}+1-j)}
  (a-\frac{M}{2}+ix)_{j-1}(a-\frac{M}{2}-ix)_{M+1-j}
  &:\text{MP}
  \end{array}\right.,
\end{align}
and those in \S\,\ref{sec:redAW} are
\begin{align}
  &\quad r_j(x^{(M+1)}_j;\bm{\lambda},M+1)\\
  &\propto e^{2ix(M+2-2j)}\times\left\{
  \begin{array}{l}
  (q^{\frac12(\alpha+\frac12)}q^{-\frac{M}{2}}e^{ix},
  -q^{\frac12(\beta+\frac12)}q^{-\frac{M}{2}}e^{ix};q^{\frac12})_{2(j-1)}\\[2pt]
  \ \ \times
  (q^{\frac12(\alpha+\frac12)}q^{-\frac{M}{2}}e^{-ix},
  -q^{\frac12(\beta+\frac12)}q^{-\frac{M}{2}}e^{-ix};q^{\frac12})_{2(M+1-j)}
  \qquad:\text{c$q$J}\\[4pt]
  \prod_{k=1}^m(a_kq^{-\frac{M}{2}}e^{ix};q)_{j-1}
  (a_kq^{-\frac{M}{2}}e^{-ix};q)_{M+1-j}
  \qquad:m=3,2,1,0\\[4pt]
  (q^{\frac12(\alpha+\frac12)}q^{-\frac{M}{2}}e^{ix};q^{\frac12})_{2(j-1)}
  (q^{\frac12(\alpha+\frac12)}q^{-\frac{M}{2}}e^{-ix};q^{\frac12})_{2(M+1-j)}
  \ \ :\text{c$q$L}
  \end{array}\right..\nonumber
\end{align}
For all these reduced cases, \eqref{eigXi1}--\eqref{VXi},
Propositions \ref{dual},\ref{polywron} and
\eqref{poldual1}--\eqref{poldual2} hold.

For example, \eqref{detiden} with $\mathcal{D}=\{\text{v}\}$ and
$N=\text{v}$ for the cases in \S\,\ref{sec:redAW} gives
\begin{equation}
  \check{P}_{\text{v}}\bigl(x;\mathfrak{t}(\bm{\lambda}),q^{-1}\bigr)
  \propto\varphi_{\text{v}}(x)^{-1}
  \text{W}_{\gamma}[\check{P}_1,\check{P}_2,\ldots,\check{P}_{\text{v}}]
  \bigl(x;\bm{\lambda}-(\text{v}+1)\bm{\delta}\bigr),
\end{equation}
which expresses a ``$q^{-1}$-polynomial'' in terms of ``$q$-polynomials''
as in \eqref{AWqi}.

%%%%%%%%%%%%%%%%%%%%%%%%%%%%%%%%%%%%%%%%%%%%%%%%%%%%%%%%%%%%%%%
%                                                             %
%  5. Summary and Comments                                    %
%                                                             %
%%%%%%%%%%%%%%%%%%%%%%%%%%%%%%%%%%%%%%%%%%%%%%%%%%%%%%%%%%%%%%%
\section{Summary and Comments}
\label{sec:summary}

Within the framework of discrete quantum mechanics for the classical
orthogonal polynomials of Askey scheme with pure imaginary shifts,
the duality between the eigenstates adding and deleting Darboux
transformations is demonstrated by proper choices of pseudo virtual
state wave functions.
The duality is based on infinitely many identities connecting the
Casoratians of polynomials of {\em twisted\/} parameters with the
Casoratians of the same polynomials of {\em shifted\/} parameters.
These identities are proven for the Wilson and the Askey-Wilson polynomials
and for every member of their reduced form polynomials, {\em e.g.\/}
the continuous (dual) ($q$-)Hahn and the continuous $q$-Hermite polynomials.

Since the logics and method of deriving these identities are almost parallel
to those for the Wronskian identities of the Hermite, Laguerre and Jacobi
polynomials, we do strongly believe that similar identities could be derived
for the classical orthogonal polynomials with real shifts, {\em e.g.\/}
the ($q$-)Racah polynomials and their reduced form polynomials.
These identities could be considered as manifestation of the characteristic
properties of the classical orthogonal polynomials, {\em i.e.\/} the
{\em forward and backward shift relations or shape invariance\/} and the
discrete symmetries. To the best of our knowledge, the discrete symmetries
for Group (B) polynomials \S\,\ref{sec:pseudoB}, which involve $q\to q^{-1}$
have not been discussed before.

The above mentioned duality itself requires  proper setting of discrete
quantum mechanics and thus valid only in a certain restricted domain of the
parameters. The Casoratian identities, \eqref{poldual1}--\eqref{poldual2},
\eqref{detiden}--\eqref{genwronide}, in contrast, are purely algebraic
relations and they are valid without any restrictions on the parameters
or the coordinates.

The multi-indexed Wilson and Askey-Wilson orthogonal polynomials are
labeled by the multi-index $\mathcal{D}$, but different multi-index
sets may give the same multi-indexed polynomials, {\em e.g.\/} eq.(3.61) in
\cite{os27}. The proposition \ref{polywron} gives its generalisation.
By applying the twist based on the type $\II$ discrete symmetry to
\eqref{detiden}, the l.h.s becomes the denominator polynomial with multiple
type $\I$ virtual state deletion and the r.h.s. becomes that of type $\II$.

Let us mention some recent works related to the solvable deformations
of classical orthogonal polynomials \cite{gomez}--\cite{quesne4}.
After completing this work, two preprints discussing similar subjects appeared
\cite{duran}.

%%%%%%%%%%%%%%%%%%%%%%%%%%%%%%%%%%%%%%%%%%%%%%%%%%%%%%%%%%%%%%%
%                                                             %
%  Acknowledgments                                            %
%                                                             %
%%%%%%%%%%%%%%%%%%%%%%%%%%%%%%%%%%%%%%%%%%%%%%%%%%%%%%%%%%%%%%%
\section*{Acknowledgements}
We thank Richard Askey for his illuminating works, which have stimulated
our imagination.
R.\,S. thanks Pei-Ming Ho, Jen-Chi Lee and Choon-Lin Ho for the hospitality
at National Center for Theoretical Sciences (North), National Taiwan University.
S.\,O. and R.\,S. are supported in part by Grant-in-Aid for Scientific Research
from the Ministry of Education, Culture, Sports, Science and Technology
(MEXT), No.25400395 and No.22540186, respectively.

\goodbreak
%%%%%%%%%%%%%%%%%%%%%%%%%%%%%%%%%%%%%%%%%%%%%%%%%%%%%%%%%%%%%%%
%                                                             %
%  References                                                 %
%                                                             %
%%%%%%%%%%%%%%%%%%%%%%%%%%%%%%%%%%%%%%%%%%%%%%%%%%%%%%%%%%%%%%%

\goodbreak

\end{document}